\documentstyle{mn}
\input{epsf}
\def\ie{{\it i.e. }}
\def\eg{{\it e.g. }}
\def\etal{{\it et al. }}
\def\deg{\ifmmode^\circ\else$^\circ$\fi}
\def\ltsima{$\; \buildrel < \over \sim \;$}
\def\simlt{\lower.5ex\hbox{\ltsima}}
\def\gtsima{$\; \buildrel > \over \sim \;$}
\def\simgt{\lower.5ex\hbox{\gtsima}}
\def\hmpc{h^{-1}{\rm Mpc}}
\def\kms{{\rm km}{\rm s}^{-1}}
\def\kmsMpc{{\rm km}{\rm s}^{-1}{\rm Mpc}^{-1}}
\def\d    {{ \rm d}}

\begin{document}

\title[Disc galaxy formation]{The Formation of Disc Galaxies}

\author[M.L. Weil, V.R. Eke and G. Efstathiou]{
M.L. Weil, $^{1,2}$, V.R. Eke, $^{1,3}$ and G. Efstathiou$^{1,3}$ \\
$^1$Department of Physics, Oxford University, Astrophysics Building,
       Keble Road, Oxford OX1 3RH\\
$^2$Astronomy Department, Columbia University, New York, NY\\
$^3$Institute of Astronomy, Madingley Road, Cambridge CB3 OHA.}

\maketitle

\begin{abstract}

We investigate the influence of the cooling epoch on the formation of
galaxies in a cold dark matter dominated universe.  Isolated haloes,
with circular speeds typical of spiral galaxies, have been selected
from a low resolution numerical simulation for re-simulation at higher
resolution.  Initial conditions for each halo consist of a high
resolution region containing dark matter and gas, and a nested
hierarchy of particles representing the mass distribution of the
original low resolution simulation.  These initial conditions are
evolved with two smoothed particle hydrodynamics (SPH) codes, TREESPH
and GRAPESPH, so that discrepancies due to differences in evolutionary
and star formation algorithms can be analysed.  In previous SPH
simulations, strong outward transport of angular momentum has led to
the formation of disc-like systems with much smaller angular momenta
than observed in real disc galaxies. Here we investigate whether this
problem can be circumvented if feedback processes prevent disc formation
until late epochs. In some of our models, the gas is evolved
adiabatically until a specified redshift $z_{\rm cool}$, at which
point the gas is allowed to cool radiatively and star formation may
begin.  In other models, cooling is continuous throughout the
evolution but suppressed by a factor chosen to allow a discs to grow
roughly linearly with time. The results of varying the cooling epoch
for each of five different haloes are analysed.  When cooling and star
formation are initiated at redshift $z_{\rm cool}=4$, stellar discs
are destroyed during merger events and we observe similar catastrophic
transport of angular momentum as seen in previous work.  With cooling
suppressed until $z_{\rm cool}=1$, discs can form by the present day
with angular momenta comparable to those of observed disc galaxies. We
conclude that feedback processes, which prevent gas from collapsing
until late epochs, are an essential ingredient in disc galaxy
formation.

\end{abstract}

\section{Introduction}\label{sec:intro}

Cosmological N-body simulations have shown that the evolution of small
density perturbations in an expanding universe can lead to a
distribution of matter very similar to the observed large-scale
structure (see {\it e.g.} Jenkins \etal 1998).  Purely N-body
simulations have also led to an understanding of the evolution of
dark matter  on the scale of individual galaxies. For example,
simulations of cold dark matter (CDM) models have shown how dark
matter haloes form by the hierarchical merging of sub-units (Frenk \etal 1988),
acquire angular momentum via tidal torques (Barnes and Efstathiou 1987),
and develop radially-averaged density profiles that satisfy simple
scaling relations (Navarro, Frenk \& White 1997).

 
Gas-dynamical simulations of the formation of individual galaxies
within dark matter haloes have been much less successful. In
simulations without star formation (\eg Navarro \& Steinmetz 1997),
the gas component can shock and dissipate energy (unlike the
collisionless halo material) forming thin discs within the dominant
halo sub-clumps.  The gas discs may be disrupted in subsequent
mergers, but the efficient conversion of kinetic energy to thermal
energy and rapid cooling allows the gas to settle into a new disc on
a timescale that is short compared to the Hubble time. The evolution
thus proceeds through a hierarchy of disc formation and merging,
leading, in the absence of star formation, to gaseous discs at the
present day.  However, during merging, angular momentum is efficiently
transported outward into the halo and the resulting gaseous discs have
angular momenta some two orders of magnitude below those of observed
spiral galaxies (Navarro \& Steinmetz 1997). Evidently, such
simulations cannot account for the formation of real disc galaxies.
We will sometimes refer to this problem as the `angular momentum
catastrophe'.

On the other hand, in simulations in which efficient star formation is
included, stars form within dark matter sub-clumps at early times.
These largely stellar sub-units then merge as the system evolves,
losing angular momentum and  forming a hot spheroidal component by the
present day.  The final objects in such simulations are
elliptical-like with no strong disc structure (\eg Katz 1992).

The failure to produce realistic disc galaxies in numerical
simulations contrasts with simple analytical models of disc galaxy
formation.  These models, first developed by Fall and Efstathiou
(1980) and Gunn (1982), show that the sizes and collapse times of disc
galaxies can be understood wihin the context of a two-component
(gas and dark matter) theory of galaxy formation 
(White and Rees 1978) {\it provided the gas component conserves
its angular momentum during collapse}. More recent applications of
this type of model have been developed to explain
other properties of disc galaxies, {\it e.g.} the slope and scatter
of the Tully-Fisher (1977) relation (Mo, Mao \& White 1998) and the
distribution of disc surface brightnesses  (Dalcanton, Spergel and Summers
1997). In fact, we will show in Section 2 of this paper, that
simple arguments can be developed to show that in CDM models, galactic 
discs cannot have lost much angular momentum during collapse. There
is, therefore, an apparent inconsistency with the `angular momentum
castrophe' described above.

In this paper we suggest that this conflict might be solved
if stellar feedback processes prevent discs forming until late epochs
($z \simlt 1$--$2$). If some fraction of the protogalactic gas is
prevented from collapsing until late epochs, when galactic dark matter
haloes are evolving slowly and most of their substructure has been erased, 
then it may be possible to form discs without
the angular momentum transport characteristic of merging sub-units. Our
aim in this paper is to test this idea using SPH simulations.

We review the observed angular momenta of disc galaxies in Section 2
and compare them with the angular momenta of CDM haloes determined
from N-body simulations. We discuss also the surface brightness
distributions of disc galaxies which we relate to the large dispersion
in the angular momenta of CDM haloes.  The SPH codes that we use, and
various other numerical details such as the star formation algorithms,
are described in Section 3. The generation of initial conditions for
the SPH simulations is described in Section 4 and our main results are
presented in Section 5.

\section{The Angular Momenta of Spiral Discs} \label{sec:obs}

In this section, we review observational constraints on the angular
momenta of disc galaxies and compare the results with the angular momenta
of dark haloes in N-body simulations of a CDM model.

We first investigate the  angular momenta of haloes identified in
numerical simulations of an $\Omega=1$ scale invariant CDM model. There
have been many previous investigations of this problem, including Barnes
and Efstathiou (1987), Frenk \etal (1988), Zurek, Quinn and Salmon (1988),
and more recently by Lemson and Kauffmann 
(1997). The results
described here are based on the simulation that is used to generate the
initial conditions of the high resolution SPH simulations described in the
main part of this paper. The numerical simulation used here is run (d) of
the ensemble described by Efstathiou (1995). The simulation is of a
spatially flat $\Omega=1$ CDM model with scale-free Gaussian initial
conditions generated from the power spectrum given by equation (7) of
Efstathiou, Bond \& White (1992) with $\Gamma = \Omega h = 0.5$\footnote
{Where $h$ is the Hubble constant in units of $100\;{\rm km}{\rm s}^{-1}
{\rm Mpc}^{-1}$ and $\Omega$ is the cosmological density parameter.}. We
assume $h=0.5$ in the rest of this paper, unless explicitly stated
otherwise.  The simulation contains $N_p = 10^6$ particles within a
cubical volume of box-length $L_{\rm box} = 40$ Mpc, with the spectrum  normalized
so that the rms mass fluctuations in spheres of radius
$8 \hmpc$ is $\sigma_8 = 0.59$ at  the present day. This normalization
approximately matches that inferred from the abundances of
rich clusters of galaxies in an $\Omega=1$ universe (White, Efstathiou \& 
Frenk 1993; Eke, Cole \& Frenk 1996).  The mass per particle in the simulation
is $4.43\times 10^9 M_\odot$.

Haloes are identified by linking together particles separated by $0.2$
times the mean interparticle separation using the percolation algorithm
described by Davis \etal (1985). The mean overdensity of these haloes is
$\Delta \approx  200$, corresponding roughly to the critical overdensity defining
the virialized region of a collapsed uniform sphere.  For each halo, we
compute the angular momentum $J$, mass $M$, total energy $E$ and circular
speed $v_c$ defined by
\begin{equation}
 v_c^2 = G {M(<R_c) \over R_c}, \label{ge1}
\end{equation}
where $M(<R_c)$ is the mass contained within a sphere of radius $R_c$
centred at the centre of mass of the halo. We used a radius of $R_c = 0.4
{\rm Mpc}$, which is approximately the virial radius of a halo with a circular
speed of $200 \;\kms$ (see equation~\ref{ge4}).  We also compute the dimensionless
spin parameter $\lambda$ (Peebles 1969)
\begin{equation}
 \lambda = J \vert E \vert^{1/2} M^{-5/2} G^{-1}. \label{ge2}
\end{equation}

\begin{figure*}
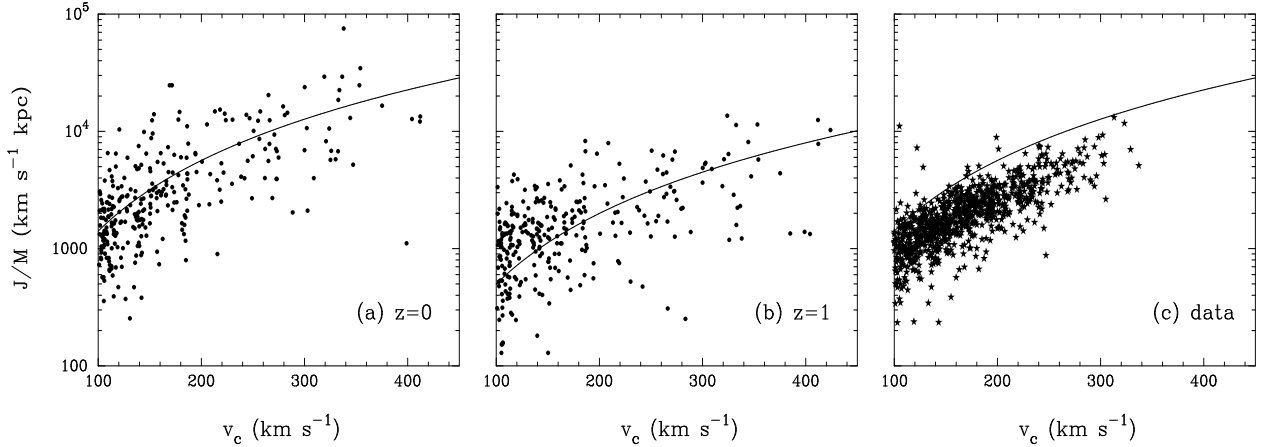

\vskip 2.7 truein

\includegraphics{pg_3.1a.ps}
\includegraphics{pg_3.1b.ps}
\includegraphics{pg_3.1c.ps}

\caption{(a) The circles show the specific angular momenta of
haloes with an overdensity $\Delta \approx 200$ identified in an
$\Omega=1$ N-body simulation of a CDM universe plotted against their
circular speeds. The solid line shows the specific angular 
momentum expected for a
halo with dimensionless spin parameter $\lambda = 0.05$ (equation~\ref{ge5});
(b) as  (a) except that the specific angular momenta have been computed
for haloes identified at $z=1$ and the circular speeds are those of the
haloes at $z=0$ within which they merge. The solid line is equation~\ref{ge5}
with $z=1$; (c) the symbols show the specific angular momenta of a sample
of disc galaxies described in the text and the solid line shows 
equation~\ref{ge5} as plotted in panel (a).}
\label{fig3.1}
\end{figure*}

Fig.~\ref{fig3.1}a
shows the specific angular momenta for haloes identified at
$z=0$ plotted against their circular speeds. This type of diagram is more
instructive than the usual plot showing specific angular momentum against
either halo or disc mass (Fall 1983, Navarro \& Steinmetz 1997), because
the circular speed of a collapsed, centrifugally supported disc is likely
to be closely similar (to within 50\%) of the halo circular speeds
(see Section~\ref{sec:sims}, and Mo \etal 1998).
With a diagram such as Fig.~\ref{fig3.1}, the circular speeds
measured for real disc systems can be used as indicators of the circular
speeds, and typical specific angular momenta, of the haloes in which they
formed.

For a halo with density profile
\begin{equation}
 \rho_H(r)  = {v_c^2 \over 4 \pi G r^2}, \label{ge3}
\end{equation}
the virial radius is given by
\begin{eqnarray}
&  R_H  & =  \left ( {2 \over \Delta} \right)^{1/2}
{v_c \over H_0} (1 + z)^{-3/2}   \nonumber \\
& \;&= 0.3 (1+z)^{-3/2} v_{300} (\Delta_{200})^{-1/2} h^{-1} {\rm Mpc},
\label{ge4}
\end{eqnarray}
where we have assumed $\Omega=1$ and the notation $v_{300} = v_c/300
\kms$, $\Delta_{200} = \Delta/200$,  {\it etc}. If we assume that the halo density
profile truncates at $r = R_H$, we can compute the specific angular
momentum from the spin parameter $\lambda$
\begin{eqnarray}
  j(R_H)  &= & \sqrt 2 \lambda v_c R_H  \nonumber \\
\lefteqn{\approx 6.5 \times 10^{3} (1 +z)^{-3/2} 
\lambda_{0.05} v_{300}^2 h^{-1} \; {\rm km} {\rm s}^{-1} {\rm kpc}^{-1}.} 
\label{ge5}
\end{eqnarray}
where we have chosen $\lambda = 0.05$, since this is approximately the
median value of $\lambda$ measured for haloes in CDM models (Barnes \&
Efstathiou 1987). Equation~\ref{ge5} is shown as the solid line in 
Fig.~\ref{fig3.1}
and provides an excellent approximation to the median specific angular
momentum as a function of circular speed.

The results plotted in Fig.~\ref{fig3.1}a
probably provide a conservative upper limit to the
angular momenta of disc galaxies.
The angular momentum of the disc component will depend on the
epoch at which the gas that forms the disc can cool, and on its radial
extent. The specific angular momentum of discs will be lower than shown in
Fig.~\ref{fig3.1}a if either the disc gas collapses from well within the virial
radius $R_H$, or if the disc collapses at high redshift. For example,
Fig.~\ref{fig3.1}b shows the specific angular momenta of haloes identified at
$z=1$ plotted against the circular speed of the halo at $z=0$ within which
they merge. The solid line in the plot shows equation~\ref{ge5}. The specific
angular momenta of haloes at $z=1$ are typically three times smaller than
at $z=0$, because the virial radius at $z=1$ is about three times smaller
than at $z=0$ and haloes typically display a nearly constant rotation
speed with radius (Frenk \etal 1988).

The third panel in Fig.~\ref{fig3.1} shows the specific angular momenta
for a large sample of disc galaxies  plotted against their circular
speeds. The photometric parameters are from Byun \& Freeman (unpublished).
These authors have analysed I-band CCD images of galaxies from 
the Mathewson, Ford \& Buchhorn (1992) sample,
using a two-dimensional profile fitting algorithm
described by Byun \& Freeman (1995). Each image has 
been decomposed into $r^{1/4}$-law bulge and 
exponential disc components, providing
estimates of the exponential disc
scale length $\alpha^{-1}$, central surface brightness $I_0$,
bulge-to-disc ratio B/D, inclination angle and I-band 
apparent magnitude. We use estimates of the circular speeds $v_c$
given in Mathewson \etal (1992), corrected for inclination, to compute
the specific angular momenta for a centrifugally supported
exponential disc with a flat rotation curve.
\begin{equation}
 {J \over M} = 2 v_c \alpha^{-1}. \label{ge6}
\end{equation}
We assume $H_0 = 50\; \kmsMpc$ to compute distances and include only 
galaxies with distances $> 10$ Mpc and bulge-to-disc
ratios $B/D < 0.3$, to reduce errors arising from inaccurate distances
and poorly determined disc profile parameters. These selection criteria
reduce the number of galaxies from $1355$ in Byun \& Freeman's sample
to $1036$.
The solid line in Fig.~\ref{fig3.1}c shows equation~\ref{ge5} with $z=0$
(\ie as plotted in Fig.~\ref{fig3.1}a). This line lies about three times
higher than the median defined by the data points. This is encouraging
because it shows that the angular momentum of a typical disc galaxy
is similar to the angular momentum within the virialized region of a 
present day CDM halo of similar circular speed.

\begin{figure*}
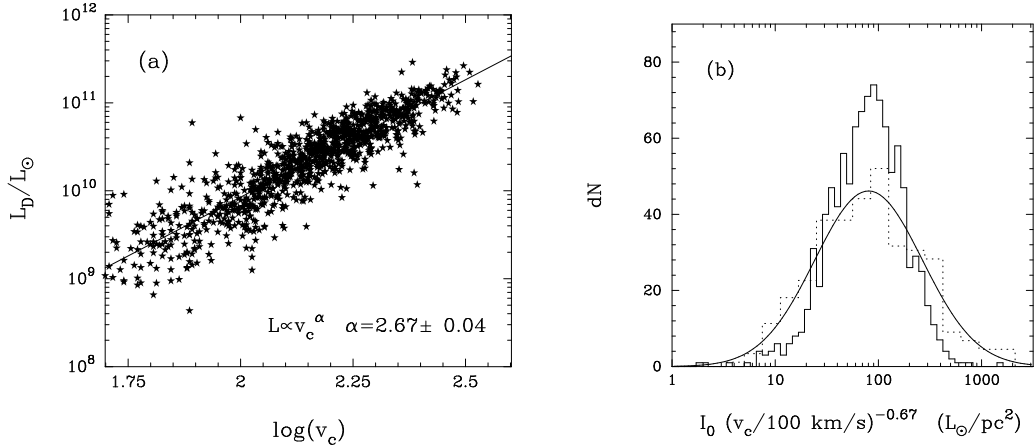


\vskip 2.8 truein

\includegraphics{pg_3.2a.ps}
\includegraphics{pg_3.2b.ps}

\caption{ (a) shows the disc luminosity of the Byun \& Freeman
sample plotted against circular speed $v_c$.  The line shows a least
squares power law fit $L_c \propto v_c^\alpha$ with $\alpha = 2.67$.
(b) compares the distribution of $I_0 v_c^{(2-\alpha)}$ for the Byun
\& Freeman sample (solid histogram) with the distribution of $1/\lambda^2$ 
determined for haloes in the CDM simulation, scaled to have approximately 
the same mean as the observations, (dotted histogram) and the best-fitting 
lognormal distribution for $\lambda$ (solid line). If the disc component
conserves angular momentum during collapse then $I_0 v_c^{(2-\alpha)}$ should
be distributed as $1/\lambda^2$.}
\label{fig3.2}
\end{figure*}

Fig.~\ref{fig3.1} also shows that the scatter in the specific angular momenta
of real disc galaxies is comparable to the scatter of the specific
angular momenta measured for N-body haloes. This can be demonstrated using the
following simple argument (Efstathiou \& Barnes 1984). Assume that a fraction
$f_D$ of the baryonic mass within the virial radius of a halo collapses
to form a disc. The luminosity of the disc is 
\begin{equation}
 L_D  =  f_D M_H \left ( {L \over M} \right)_D = 
f_D {v_c^3  \over H_0G} 
\left ( {L \over M} \right)_D \left ( {2\over \Delta} \right)^{1/2} \propto v_c^\alpha,
\label{ge7}
\end{equation}
where the expression for $M_H$ has been substituted assuming that 
$\Omega=1$ and $z=0$.
The proportionality expresses the tight Tully-Fisher (1977) 
correlation between luminosity and circular speed observed for real 
disc systems. Evidently, the (uncertain) cooling and feedback processes which
determine $f_D$ must be tightly correlated with circular speed to 
reproduce the Tully-Fisher relation. The correlation between disc
luminosity and circular speed for the Byun-Freeman sample is shown
in  Fig.~\ref{fig3.2}a. A least squares fit to the data points gives
an exponent $\alpha = 2.67 \pm 0.04$. If we equate the specific 
angular momentum of the disc to $j(R_H)$ as given by equation~\ref{ge5} 
at fixed $v_c$, then the collapse factor of the disc
is related to the spin parameter of the halo according to
\begin{equation}
 \alpha R_H = {\sqrt 2 \over \lambda},
\label{ge8}
\end{equation}
(Fall \& Efstathiou 1980). If the gas conserves its angular
momentum during collapse, then a disc that forms in a halo with
a low value of $\lambda$ will collapse by a large factor so
ending up with a high mean surface mass density; likewise
a disc that forms in a  halo with a high value of $\lambda$ will 
end up with a low mean surface mass density.  The large scatter
in $\lambda$ from tidal torques should therefore be reflected in
the spread of surface mass densities and surface brightnesses of
disc systems. The disc luminosity of an exponential disc
is related to the central surface brightness, $I_0$, according to
$L_D =2 \pi I_0 \alpha^{-2}$, thus from equations~\ref{ge7}
and (\ref{ge8}),
\begin{equation}
 I_0 v_c^{(2-\alpha)} \propto  {1 \over \lambda^2}.
\label{ge9}
\end{equation}

Fig.~\ref{fig3.2}b shows the distribution of $I_0 v_c^{-0.67}$
for the Byun-Freeman sample, compared to the distribution
of $1/\lambda^2$ for haloes in the N-body simulations. The mean
of the $1/\lambda^2$ distribution has been adjusted to match the
mean of the observational data points. The observed distribution
plotted in Fig.~\ref{fig3.2}b  peaks at about $100 \;L_\odot/{\rm pc}^2$
and has a broad distribution comparable to, but not quite as broad
as the distribution of $1/\lambda^2$. It is easy to think of 
reasons why the observed distribution might be artificially narrow,
for example, the Byun-Freeman sample could be incomplete at low
and high surface brightness, and perhaps some objects with large
collapse factors produce spheroidal systems. Indeed there is a large
literature on observational selection effects and the surface brightness
distributions of galaxies (\eg Disney 1976; Allen \& Shu 1979;
Phillipps \& Disney 1986, Dalcanton, Spergel and Summers
1997). Qualitatively, however, the above comparisons
show that the mean angular momenta of disc systems, and their scatter, fit
well with a picture in which disc galaxies form at recent epochs
by the collapse of extended gaseous atmospheres within dark haloes.
Fig.~\ref{fig3.2}b suggests that disc galaxies form within dark haloes with 
a wide range of $\lambda$ values, \ie that the angular momentum
of the halo is not a critical factor in determining whether a disc 
system forms. Within this picture, however,  the
angular momentum of the disc component must be approximately conserved
during collapse.

In the following sections, we will investigate this picture numerically
in some detail, showing that cooling and feedback processes are
crucial in determining the angular momentum of the disc component,
as well as in determining the fraction of the baryonic gas that 
can cool.

\section{Numerical Techniques}\label{sec:num}

Our cosmological simulations were performed using two different
self-consistent, 3-dimensional, N-body  codes,  
TREESPH (Hernquist \& Katz 1989) and GRAPESPH (\eg Steinmetz 1996).
These codes are capable of evolving systems containing both collisionless
material (\ie stars and dark matter) and gas.

Smoothed particle hydrodynamics, in which gas is partitioned
into fluid elements, is used to model hydrodynamical interactions
between gas particles (Lucy 1977, Gingold \& Monaghan 1977).  The
self-gravitating fluid elements are represented as particles that
evolve according to Lagrangian hydrodynamic conservation laws
which take into account local effects arising from pressure gradients
and viscosity.  The equation of state is that for an ideal gas, 
$P=(\gamma -1)\rho u$ where $\gamma=5/3$, $\rho$ is the gas density, 
and $u$ is the specific thermal energy.
As shocks can arise in the gas, an artificial viscosity is
required to maintain numerical stability.  The form of artificial 
viscosity used
in both codes is similar to that used by 
Navarro \& Steinmetz (1997), {\it i.e.} modified by the curl
of the velocity field to reduce the shear component.

Radiative cooling and star formation are included in the codes.
The primordial gas is assumed to have  a hydrogen fraction
of $X=0.76$ and a helium fraction of $Y=0.24$ by weight.  
Cooling is computed using the 
ionization, recombination, and cooling rates from Black (1981, Table 3) with
the modifications suggested by Cen (1992) for 
temperatures $T \simgt 10^5$.  
Radiative cooling does not 
proceed below a minimum temperature $T_{min}=10,000$K.
The star formation is implemented differently in the two codes and will be
described in Sections 3.1 and 3.2.

\subsection{TREESPH}

TREESPH computes gravitational forces between particles
using a hierarchical tree algorithm 
that groups distant particles into nested cells.
The potentials of the cells are then approximated with a multipole series 
truncated at the quadrupole.  When computing the force on a particle, 
TREESPH calculates the ratio of the current cell size to
the distance between the cell and the particle.
For values $\leq \theta$, a specified tolerance parameter, 
the force from the cell is treated as a whole;
otherwise, the next cellular subdivision is considered.
A spline kernel is used to soften the gravitational potential in order
to reduce two-body relaxation.  

In TREESPH, individual smoothing lengths, $h_i$, are
calculated for each gas particle and updated to ensure that
$N_{smooth}=25-45$ neighbours are contained within a sphere of radius
$2h_i$.  In addition to individual smoothing lengths,
individual timesteps are permitted for each particle.  
The largest time step for these TREESPH simulations 
is $\Delta t = 6.24 \times 10^6$ years; the smallest number of timesteps 
in which a particle can reach $z=0$ is $N_{step}=2000$.  Smaller timesteps
are chosen for particles in dense regions, using the criterion 
$a_i v_i \Delta t_i \leq e_{tol} E_i$ where $a_i$ and $v_i$ are the
acceleration and velocity of the particle, and $e_{tol}=0.1$ determines the 
fractional accuracy of the integrations.  $E_i$ is the maximum of the
mean specific kinetic energy and one sixth of the mean specific potential
energy of the system (Ewell 1988).
Particles in the
highest density areas are allowed to reduce their timesteps by up to
a factor of eight.  

Thermal energy is computed from 
\begin{equation}
\begin{array}{ll}
	\frac{\d u_i}{\d t}=&\sum_j m_j ({P_i \over \rho_i} 
	+ {1 \over 2} \Pi_{ij}) {\bf v}_{ij} \cdot {1 \over 2} 
	[{\bf \nabla}_i W(r_{ij},h_i) + \\
	&{\bf \nabla}_i W(r_{ij},h_j)]-\frac{\Lambda_i}{\rho_i}\\
	\end{array}
\end{equation}
where the first term is due to adiabatic processes, the second term 
accounts for the viscosity, and the third term includes the remaining
nonadiabatic radiative cooling.  
$W$ is the SPH smoothing kernel, which is
implemented in the gather-scatter formulation of Hernquist \& Katz (1989).
A semi-implicit method of integrating the energy equation is employed.
An explicit integration using a timestep that is half of the shortest SPH
particle timestep is performed on the adiabatic terms whereas 
an implicit integration
is performed on the nonadiabatic thermal heating and cooling processes.
In high density regions, the cooling time step can become so short that it 
is computationally impractical to employ it.  
Thus, in order to avoid numerical errors in the
integration of the thermal energy, the cooling rate is damped so that a
particle can lose no more than half its thermal energy during a cooling
time step (Katz \& Gunn 1991).  

Star formation, in which gas particles which meet the formation criteria
are allowed to form a `star' wholly and instantaneously, has 
been included in TREESPH.  The criteria for star formation are similar to
those of Katz (1992).  First, gas particles in collapsing regions are found,
using $\nabla \cdot {\bf v_i} < 0$ as the condition  for the existence of
a convergent flow.  Second, the flow is tested for Jeans instability.
The particle is locally Jeans unstable if the sound crossing time is shorter
than the dynamical time.  Each gas particle in a converging flow with
\begin{equation}
{h_i \over c_i} > {1 \over \sqrt{4 \pi G \rho_i}},
\end{equation}
where $c_i$ is the local sound speed, is thus a candidate for star formation.  
However, regions in which the gas density has decreased drastically 
during previous star formation will not contract as readily as physically 
expected due to gravitational softening.  Because gas particles in these 
high star formation regions will otherwise be prevented from forming stars 
through inability to meet the Jeans instability criterion, that criterion 
is not applied to gas in softened regions with 
\begin{equation}
K_s ({\pi \epsilon_i \over \sqrt{2}})^2 {\mu m_H G \over k T_i} \rho_i >10,
\end{equation}
where $K_s=0.89553$ is a constant derived from a calculation of the Jeans mass
(Katz 1992), $\epsilon_i$ is the softening length, $m_H$ is the
mass of a hydrogen atom, $k$ is the Boltzmann constant, 
and $T_i$ and $\rho_i$ are the gas temperature and density.  
Third, if the cooling time, $t_c=u_i(du_i/dt)^{-1}$, is shorter than 
the local dynamical time, $t_d=(4 \pi G \rho_i)^{-1/2}$, then gas pressure will
not inhibit the collapse of the gas particle and it may form a star.
However, because radiative cooling is cut off at $T_{min}=10,000$K, gas 
particles at low temperatures have long cooling times.  In order to
avoid an artificial cessation of star formation in low temperature regions,
gas particles with $T<30,000$K which otherwise meet the criteria 
are allowed to form stars with 
a cooling time set to the dynamical time $t_c=t_d$. 
The local rate of star formation is 
\begin{equation}
{d \rho_* \over dt}= {-d \rho_g  \over dt} = {c_* \rho_g \over t_c}
\label{sfrts}
\end{equation} 
where $c_*=0.1$ is a dimensionless star formation parameter.
The probability that a gas particle will form a 
star in time $\Delta t$ is 
\begin{equation}
p_*=1 - e^{-c_* \Delta_t /t_c}.
\end{equation}
A random number is generated for each star candidate;
if it is less than $p_*$, the gas particle loses its SPH characteristics and
becomes a collisionless star particle.

For these cosmological simulations, TREESPH evolves the haloes from
the initial redshift to $z=0$ using 
physical coordinates with isolated boundary conditions.

\subsection{GRAPESPH}

The version of GRAPESPH we employ has evolved from the TREESPH
code of Navarro \& White (1993; hereafter NW). Rather than using the mutual 
nearest-neighbour binary tree (Benz \etal 1990) of the original implementation
in order to calculate gravitational forces and find neighbours for the SPH
force evaluation, the special-purpose GRAPE-3Af hardware (Sugimoto \etal 
1990) has
been harnessed to perform these tasks. This machine carries out a parallelised
direct summation over all particles and very rapidly returns the forces and
neighbour lists. All hydrodynamical forces are then calculated on the host
workstation. The adaptations which convert a
TREESPH code into a GRAPESPH code have been described in detail by 
Steinmetz (1996).

Both individual timesteps and individual gas particle smoothing
lengths were used. As the GRAPE only returns `gather' neighbours, that
is those within distance $h_i$ of gas particle $i$, the neighbour
search radius is taken to be $\sqrt{1.5}$ times the smoothing
length. This should ensure that the actual number of mutual
neighbours, which satisfy $h_i+h_j<$ twice the particle separation,
lies within the range $25-45$.

The gravitational forces are calculated assuming a Plummer softening law. For
each particle, the total acceleration due to gravity is found in two steps. 
First, the acceleration arising from all gas particles is found, and then that
resulting from dark matter and stars is incremented. A different gravitational
softening parameter is used in these two separate force evaluations.

A simplified version of the star formation algorithm described by 
NW was adopted. Any gas particle in a sufficiently 
high density, collapsing region 
is considered to be a potential star-forming particle.
These two constraints can be expressed as
\begin{equation}
\rho_{\rm i} > \rho_{\rm crit*}=7\times10^{-23}~{\rm kg~m^{-3}}
\label{star1}
\end{equation}
and
\begin{equation}
{\bf \nabla}.{\bf v_{\rm i}} < 0.
\label{star2}
\end{equation}
The local dynamical time is then defined by
\begin{equation}
\tau_{\rm dyn} = \sqrt{\frac{3\pi}{16 G \rho_{\rm i}}}.
\end{equation}
If conditions (\ref{star1}) and (\ref{star2}) remain satisfied for the ensuing 
dynamical 
time then the gas particle converts entirely to a star particle. Should either
of these constraints be violated during this probationary period then the gas 
particle is no longer considered to be star-forming and it remains gaseous.

\subsection{Contrasting the Codes}

The softening of the gravitational interaction between two particles 
differs in the two codes.  In GRAPESPH, the potential is of the
form corresponding to a Plummer density profile,
$\Phi \propto (r^2 + \epsilon^2)^{-1/2}$.   
However, the acceleration calculated using a Plummer model softening 
converges to the Kepler value relatively slowly. 
In codes like TREESPH, in which distant particles
are assigned to cells and their potentials are represented by quadrupole
expansions, the cells are assumed to be point-like rather than extended.
Thus, a softening in which the acceleration quickly converges to the Kepler 
form is advantageous.
In TREESPH, the gravitational potential
is softened using the spherically symmetric spline kernel suggested by
Monaghan \& Lattanzio (1985) because it has compact support and is identical
to the Kepler form for $r \geq 2\epsilon$.  
For the effective softening to be similar in 
both codes, the TREESPH softening lengths must have values approximately
twice those of the GRAPESPH softening lengths.

The star formation algorithms employed within the two codes derive
from Katz (1992) and NW.  They are similar in requiring that a gas
particle be able to form a star only if it is in a converging flow.
However, other criteria differ.  This version of GRAPESPH, unlike
TREESPH and NW, does not use the Jeans instability criterion for
creating star particles nor does it define a star formation rate.
Like NW, GRAPESPH defines a critical density, above which the cooling
time is less than the dynamical time.  The two GRAPESPH criteria are
applied to each particle when it is advanced.  The criteria are
monitored over the dynamical time for each `successful' gas particle
after which, if they are not violated, a star is formed.  In general,
the delay is $\simlt10^7$ years.  In TREESPH, the criteria are checked
at each system timestep of $\Delta t = 6.24 \times 10^6$ years; a
random number of gas particles which meet the star-forming criteria
are chosen to form stars immediately.  The effect of these differences
is that the GRAPESPH simulations produce more stars between $z=1$ and
$0.8$ while, afterwards, the star formation rate in the TREESPH
simulations is higher.  By $z=0$, the number of GRAPESPH stars is some
$75-90\%$ of the number of TREESPH stars.  This difference is due to
the additional TREESPH formation criteria which allow stars to
continue to form in regions of high star formation where the gas
density has decreased due to the transformation of gas particles to
stellar particles.  Qualitatively, the GRAPESPH criteria act to mimic
larger supernova feedback effects in regions of high star formation.

\section{Initial Conditions}\label{sec:ics}

The initial conditions for our simulations are generated using a
nested hierarchy algorithm similar to that described by Katz \& White
(1993). The idea is to select haloes from the dissipationless CDM
simulation described in Section~\ref{sec:obs}, and then to re-simulate
them at higher mass resolution using the SPH codes of the previous
section. This technique has been used by 
Navarro, Frenk \& White (1995b) and Eke, Navarro \& Frenk (1998), 
in SPH simulations of clusters of galaxies and by
Navarro \& Benz (1991), 
Quinn, Katz \& Efstathiou (1996), Navarro \& Steinmetz (1997) and
others in SPH simulations of galaxy formation.

\begin{figure*}
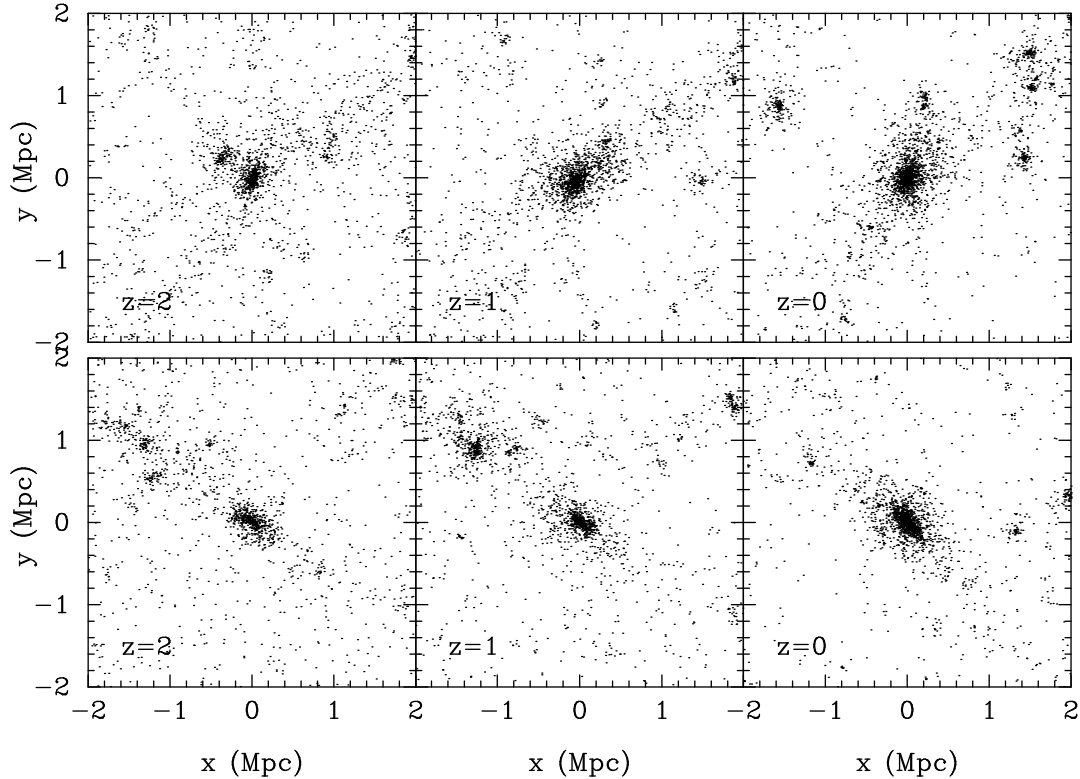


\noindent
\vskip 4.2 truein

\includegraphics{pg_4.1a.ps}
\includegraphics{pg_4.1b.ps}

\caption{The evolution of two haloes from the dissipationless
N-body simulation described in Section~\ref{sec:obs}.
x-y projections are shown of the particles located within a box of comoving
length $4 \; {\rm Mpc}$ centred on the dominant subclump at $z=2$,
$z=1$ and $z=0$. The upper panel shows the evolution of Halo
2 listed in Table 1, with a final circular speed of $v_{\rm c}=249~\kms$. 
The lower panel shows the evolution of
a halo with a nearly identical final circular speed, $v_c = 247~\kms$,
that we did not simulate with the SPH codes because it evolves
by the merger of two comparable mass components at $z< 1$.}
\label{fig4.1}
\end{figure*}

\subsection{Selection of the Haloes}

Haloes with a density contrast $\Delta \approx 200$ were located at
$z=0$ in the N-body simulation using the group finding algorithm
described in Section~\ref{sec:obs}. The haloes were ordered by
circular speed. We then visually examined plots of the halo particles
at various output times and selected five examples for simulation with
the SPH codes. The haloes were chosen to have a range of circular
speeds between $300\; \kms$ and $150 \;\kms$, to be far from much more
massive haloes, and not to have merged with a comparable mass system
since $z=1$. Parameters for the 5 chosen haloes are given in Table
1. The first condition selects objects with circular speeds in the
range observed for normal luminous disc systems. The second criterion
eliminates satellite systems near the outer parts of massive systems,
for which our tree codes and hierarchical initial conditions algorithm
are ill-suited. The third criterion eliminates haloes that undergo
major merger events at low redshifts and so preferentially selects
regular objects that grow by the accretion of smaller systems at low
redshifts.  Nevertheless, even with this selection criterion, the
typical halo mass within an overdensity of $\Delta = 200$ nearly doubles
between a redshift of $1$ and $0$ (see Table 1).

\bigskip

\centerline{\bf Table 1: 
Parameters for the five haloes}
\centerline{\bf chosen for SPH simulation}
\begin{center}
\begin{tabular}{|cccccc|} \hline
      &               & \multicolumn{2}{c}{$M_H \times 10^{11} M_\odot$}
& \\
Halo  &  $v_c$ (km/s) & $z=0$ & $z=1$ &  $\lambda_H$ & $L_{hr}$ (Mpc)\\ \hline
$1$   & $265$ &  $70.5$ & $35.2$ & $0.050$ &  $6.00$\\       
$2$   & $249$ &  $57.7$ & $38.8$ & $0.020$ &  $6.08$\\
$3$   & $176$ &  $25.1$ & $14.2$ & $0.031$ &  $4.92$\\   
$4$   & $160$ &  $21.5$ & $11.8$ & $0.040$ &  $5.64$\\      
$5$   & $157$ &  $20.1$ & $16.7$ & $0.032$ &  $4.84$\\
 \hline
\end{tabular}
\end{center}
{\it Notes to Table 1:} The first column gives the halo number, used
throughout this paper. The second column gives the circular speed computed
from equation~\ref{ge1}. 
The third column gives the mass of the halo at $z=0$,
and the fourth column gives the mass of the dominant subclump at $z=1$
identified by running the group finding algorithm with a link parameter
$b=0.2$. The fifth column gives the $\lambda$ parameter of the halo at
$z=0$. The final column gives the comoving size of the box used to
generate high resolution initial conditions as described in 
Section~\ref{ssec:ics}.

\bigskip

The upper panel in Fig.~\ref{fig4.1} shows an x-y projection of Halo
$2$ at $z=2$, $z=1$ and $z=0$. The lower panel shows the
evolution of a halo with almost identical circular speed
 that we decided not to simulate because
it evolves by the merger of two comparable mass subunits at
$z < 1$. The selection criteria that we have imposed will
bias statistics of the frequency of disc galaxy
formation in CDM models, and this must be borne in mind when
interpreting some of the results presented in the next two 
Sections. In particular, the last column of Table 1 shows that
our selection criteria bias against haloes with high values of
$\lambda$. This is because haloes with high $\lambda$ often
contain two distinct subclumps, or are near much more massive
systems.

\subsection{Generation of High Resolution Initial Conditions}\label{ssec:ics}

 The particles within a sphere of radius $400$ kpc around the
centre of mass of the halo at $z=0$ were traced back to the initial
conditions. The smallest cubical box containing all of these
particles was located, defining the region to be resampled
at higher mass resolution. The comoving box sizes of the high
resolution regions for each halo are listed in the final
column of Table 1; typically a box of comoving size
$5$--$6$ {\rm Mpc} is required. Having located the size and
location of the high resolution region, initial conditions
for the SPH simulations were generated as follows.

\noindent
[1] Distribute $N_{hr}^3$ particles on a regular lattice
within the high resolution box of size $L_{hr}$.

\noindent
[2] Generate displacements of these particle positions using
the algorithm of Efstathiou \etal (1985) with the identical
amplitudes and phases used to generate the initial conditions
of the low resolution simulation, but with the initial power
spectrum sharply truncated at the Nyquist frequency of the
low resolution particle grid $k_{n} = \pi N_p/L_{\rm box}$.

\noindent
[3] Generate additional displacements to add in the missing
small-scale power between the Nyquist frequencies of the 
low resolution particle grid and the high resolution particle
grid $k_{n^\prime} = \pi N_{hr}/L_{hr}$ from the appropriate
CDM power spectrum.

\noindent
[4] Add $M$ nested layers of massive particles to represent
the tidal field of the surrounding matter. The particles in these
outer layers are arranged on lattices such that the lattice spacing 
increases by factors of two outwards from the high resolution region
and particle masses increase by factors of eight, as described by
Katz \& White (1993, see their Fig. 1). Five layers were
chosen to match exactly the total
mass, $M_{\rm box}$, contained in the low resolution N-body simulation.

\noindent
[5] Generate displacements for these outer particles as in [2]
above, but with the power spectrum  sharply truncated below the
Nyquist frequency of each layer. This last step prevents aliasing 
of small-scale power from affecting
the representation of the tidal field.

For the simulations described in the next section, we used 
a high resolution region represented by $27^3$ particles.
The outermost of the five low resolution layers of dark matter
particles was stripped off  each halo, because tests revealed that it
had a negligible effect on the evolution of the particles 
within the high resolution region.    The total mass, $M_{\rm tot}$, of the
four remaining layers and the interior high resolution particles was
calculated and the new outermost layer was truncated by removing all
particles within radius
\begin{equation}
r  =  {1 \over 2} 
\left ( {M_{\rm tot} \over M_{\rm box}} \right )^{1/3} L_{\rm box}, 
\label{rme}
\end{equation}
from the centre of the simulation.
This 
produces a spherical distribution of the outermost high mass
particles.  Each dark matter particle in the high resolution layer was
reduced in mass by 10\% and overlaid with a gas particle of mass 
$m_{gas}=0.1m_{dark}$ to produce a high resolution region containing  gas.
The masses of dark matter and gas particles in the simulations range from 
$3.6 - 7.1 \times 10^8 M_{\odot}$ and $4.0 - 7.9 \times 10^7 M_{\odot}$,
respectively.

The total number of particles is $N_{tot}=41136$, with the
low resolution layers having $N_1=1178$, $N_2=386$, $N_3=152$, and
$N_4=54$.  In the TREESPH simulations, the gas particles were offset
in position from the dark matter particles by a fraction of a
softening length to avoid numerical difficulties in
constructing a tree containing all particles.

\section{Numerical Simulations}\label{sec:sims}

\subsection{Parameters of The Simulations}

Each of the five haloes of Table 1 was simulated at high mass
resolution with TREESPH and GRAPESPH. As we will show in
subsequent sections, the softening of the gravitational force law has
a large effect on the calculations. In the TREESPH simulations, the
high resolution dark matter particles are assigned gravitational
softening lengths of $\epsilon_{dark}=10.4$ kpc.
The gas and star particles in TREESPH have
$\epsilon_{gas}=\epsilon_{stars}=2.6$ kpc to allow better resolution
of the overdense regions in which the galaxies form.  Larger softening
lengths are adopted for the low resolution layers of high mass
particles according to $\epsilon_{layer}=
(m_{layer}/m_{dark})^{1/3}\epsilon_{dark}$.  In the
GRAPESPH simulations, the gravitational forces are computed assuming a
Plummer force-law with softening lengths of $\epsilon_{dark,star}=5$
kpc for the dark matter and star particles and $\epsilon_{gas}=1$ kpc.
With these choices, the softened force laws for the gas and high
resolution dark matter particles are similar in the two codes.
However, the softening adopted for the stars in GRAPESPH is about twice
as large as the softening adopted for the star particles in
TREESPH. This is because in the GRAPESPH
code, gravitational forces for all collisionless particles (dark matter and 
stars) are computed in one single parallelized direct summation
and thus have the same softening.

Both codes employ variable timesteps. Particles in the TREESPH runs
completed the evolution from the initial redshift of $z=7.4$ to $z=0$
in $2000$ to $16000$ timesteps and a typical simulation with $41136$
particles required approximately $100$ hours to run on an DEC Alpha
4100 5/300. Evolution of the particles in the
GRAPESPH simulations required typically $\sim 100-40000$ steps
and $\approx 45-65$ CPU hours to run to completion from
$z=7.4$ to $z=0$ on the GRAPE machine at Edinburgh.

In most previous simulations of this type, the gas has been allowed
to cool radiatively with no other energy input except that
provided by gravitational shock-heating and, in some simulations,
a uniform photoionizing background radiation (see \eg
Quinn \etal 1996; Navarro \& Steinmetz 1997). In some simulations,
attempts have been made to mimic energy injection into the intergalactic
medium by supernovae (Katz 1992; NW). 
Broadly speaking, these have involved either injecting energy
directly as heat into the gas, or into bulk motion so reversing
the flow of gas into high density regions. As discussed by NW,
neither approach is satisfactory. Energy supplied as heat
into high density gas is radiated away so efficiently that it 
has very little feedback effect. On the other hand,
it is easy to reverse the flow of gas by injecting energy
into bulk motions, but the feedback effects are then extremely
sensitive to the specific algorithm and parameters that are
used. We believe that a more realistic description of feedback
requires the ability to model a multiphase gas component, in
which cool high density gas can coexist spatially with an outflowing
hot lower density gas component. Modeling such a multi-phase
gas medium  would require fundamental modifications to the
SPH codes described here, and so is well beyond the scope of
this paper. Some provisional attempts along these lines, using
a Eulerian gas dynamic code, are described by Yepes \etal (1997).

Nevertheless, as described in the Introduction, 
our thesis in this paper is that feedback processes are
critical in determining the angular momenta of disc systems.
The key process that we wish to model is the late infall of
gas into a dark matter halo. The numerical simulations
of Navarro \& Steinmetz (1997) described in the introduction
show that, in the absence
of any feedback, gas collapses at high redshifts
into dark matter subclumps that later merge, losing most of
their orbital and internal angular momentum to outer halo
particles. If, however, feedback processes can prevent
some significant fraction of the gas from collapsing at high 
redshifts,  then there is a possibility that this residual gas 
can collapse within a more uniform and slowly evolving halo 
to form a disc system whilst conserving its angular momentum.
Such a formation scheme for disc systems seems essential
in hierarchical clustering theories, since
the discussion of Section~\ref{sec:obs} shows that the angular momenta
of disc systems can only be understood if the gas
approximately conserves its angular momentum during
collapse.

In this paper, we have chosen a particularly simple scheme to model
feedback processes. We evolve the gas component adiabatically
until a specified redshift $z_{\rm cool}$, at which point we switch on
radiative cooling with a cooling rate $\Lambda$ as described in 
Section~\ref{sec:num}.
For most of the runs, we adopted $z_{\rm cool} = 1$. Some runs were
done with $z_{\rm cool}=4$ and $z_{\rm cool}=0.6$ to illustrate the sensitivity
of angular momentum evolution to the epoch of gas collapse. Two simulations
were done in which the cooling rate $\Lambda$ was suppressed by a 
factor,
\begin{equation}
g(z)=(1+z)^{\alpha} {\rm exp} \big (-  (z/ z_c )^2 \big ), \qquad \alpha=-1.5,
\label{gz}
\end{equation}
where $z_c$ is a parameter. The parameter $\alpha$ was fixed to
 $\alpha=-1.5$ by 
an analytical calculation of cooling within an isothermal
halo so that discs would grow roughly linearly
with time. Simulations were done with $z_c=1$, almost completely eliminating
high redshift cooling, and with $z_c=4$. Table 2 gives a complete list of
the SPH simulations that were run for this paper.

\subsection{Comparison of Codes}\label{ssec:comp}

The simulations described here are at the limit of what is possible with
present-day computers, thus it is important to understand the
limitations imposed by the computational scheme, such as limited mass
resolution, gravitational softening {\it etc}.  In this subsection, we
thus describe in some detail the differences in the evolution of
two haloes (2 and 3)  run with the two codes.

\subsubsection{Morphologies}\label{ssec:morph}

Fig.~\ref{fig:fig5.1.1} shows projections of the dark matter and
gas distributions at $z=1$ evolved with cooling suppressed (Runs 4, 5, 11,
and 12). The left-hand panels show the particle distributions evolved
with TREESPH and the right-hand panels those evolved with
GRAPESPH. Although the two codes produce dark matter and gas
distributions that are closely similar, we were surprised to see some
differences. Perhaps the most striking variations can be seen
in the relative positions and morphologies of the four dark matter
concentrations to the left of Halo 3. The structure of the central
halo itself also appears to be slightly different in the two runs.

To explore these morphological differences, and determine whether they were
caused by differences in the treatment of the gas component, we
re-evolved Halo 3 with the two codes from $z=7.4$ to $z=1$ using only
dark matter particles.  First, the haloes were run with the same
computational parameters as Runs 11 and 12. Projections at $z=1$

\bigskip

\centerline{\bf Table 2: List of Simulations}
\begin{center}
\begin{tabular}{|ccccc|} \hline
Run   & Halo & Code &  $z_{\rm cool}$ & $z_c$  \\ \hline
1     &  1   &  TREESPH  & 1 & -- \\
2     &  1   &  GRAPESPH & 1 & -- \\
3     &  1   &  TREESPH  & 4 & -- \\
4     &  2   &  TREESPH  & 1 & -- \\
5     &  2   &  GRAPESPH & 1 & -- \\
6     &  2   &  TREESPH  & 4 & -- \\
7     &  2   &  GRAPESPH & 4 & -- \\
8     &  2   &  GRAPESPH & 0.6 & -- \\
9     &  2   &  TREESPH  & -- & 1 \\
10     &  2   &  TREESPH  & -- & 4 \\
11     &  3   &  TREESPH  & 1 & -- \\
12     &  3   &  GRAPESPH & 1 & -- \\
13     &  4   &  TREESPH  & 1 & --\\
14     &  4   &  GRAPESPH  & 1 & -- \\
15     &  5   &  TREESPH  & 1 & -- \\
16     &  5   &  GRAPESPH & 1 & -- \\
17     &  5   &  TREESPH  & 4 & -- \\
18     &  5   &  TREESPH  & 0.6 & -- \\  
 \hline
\end{tabular}
\end{center}
{\it Notes to Table 2:} The first column lists the run number, the second
gives the halo number as in Table 1 and the third column lists
code used to run the simulation. The fourth column lists the redshift
$z_{\rm cool}$ at which radiative cooling was switched on. Where this is
blank, we suppressed the cooling rates by the factor given in
equation~\ref{gz} with parameter $z_c$ listed in the fifth column.

\bigskip

\begin{figure}
\centering
\centerline{\epsfxsize=10.0cm \epsfbox[20 170 350 710]
}
\caption{Projections of the dark matter and gas
components at $z=1$ for two haloes (Runs 4, 5, 11, and 12) evolved with the
TREESPH and GRAPESPH codes.}
\label{fig:fig5.1.1}
\end{figure}

\begin{figure}
\centering
\centerline{\epsfxsize=9.0cm \epsfbox[80 170 550 700]
}
\caption{Four simulations of Halo 3, using dark matter only,
plotted at $z=1$. Panels (a) and (b) show TREESPH
and GRAPESPH using the default computational parameters.
These plots are almost identical to the dark matter distributions
plotted in Fig.~\ref{fig:fig5.1.1} and show distinct differences between the
two codes, for example, in the  location of the four clumps
to the left of  the main central mass concentration. Panels
(c) and (d) show a TREESPH simulation employing a  more 
accurate force computation, and a GRAPESPH simulation using smaller
timesteps. Panels (c) and (d) are almost identical, and similar
to the default GRAPESPH simulation, showing that
force errors in TREESPH cause most of the discrepancies.
}
\label{fig:fig5.1.1ac}
\end{figure}

\noindent
 are
shown in panels (a) and (b) of Fig.~\ref{fig:fig5.1.1ac}, which can be 
compared directly
with the dark matter plots shown in Fig.~\ref{fig:fig5.1.1}.
The differences between the particle positions computed from the
two  codes are almost identical to those seen
in Fig.~\ref{fig:fig5.1.1},  implying that
they do not arise merely from differences in 
the treatment of gas. Rather, the differences must arise from errors
in the gravitational forces or the time-integration of the equations of
motion. We therefore reran these dark matter only runs using different
code parameters. For TREESPH, we improved the accuracy of the
forces  by changing the force-tolerance parameter from the
default value  $\theta = 0.8$ to $\theta=0.5$. In GRAPESPH, the force
calculation is hardwired into the GRAPE with, typically, a $\sim 2\%$ 
dispersion in the forces between particle pairs (see Okumura \etal 1993). 
Since the force accuracy cannot be adjusted in the GRAPESPH code, we
reran the simulation employing three times as many timesteps. Snapshots
from these two runs are shown in panels (c) and (d) of 
Fig.~\ref{fig:fig5.1.1ac}, and 
show clearly that force errors in TREESPH are the main cause of the 
discrepancies. For many purposes these small errors are unimportant,
but they can lead to different merger histories and to differences in
the radial profiles of some quantities sensitive to the presence
of substructure, {\it e.g.} plots of the specific angular momentum
with radius (see Figures 8 and 12 below).

\begin{figure*}
\centering
\centerline{\epsfxsize=10.0cm \epsfbox[144 210 400 620]
{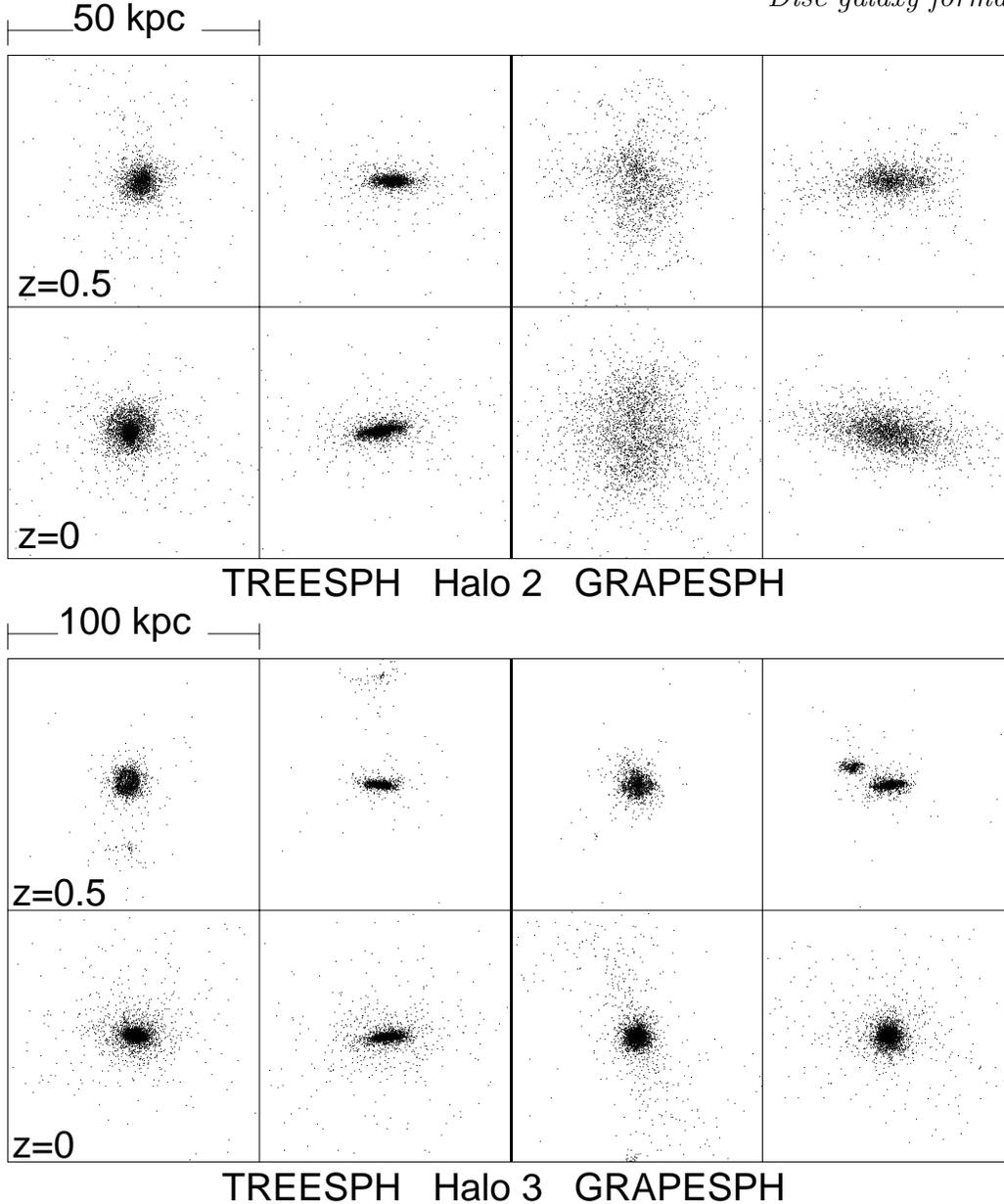}}
\caption{Stars in the Halo 2 (Runs 4 and 5)
and Halo 3 (Runs 11 and 12) simulations at $z\approx 0.5$ and $z=0$.  
Lefthand panels show the TREESPH simulations 
projected along the principal axes $X - Y$ and $X - Z$ and the
righthand panels show the  GRAPESPH simulations in the 
same projections.}
\label{fig:fig5.1.2}
\end{figure*}

The stellar condensations that form in Runs 4 and 5 (Halo 2)
and Runs 11 and 12 (Halo 3) are
shown in Fig.~\ref{fig:fig5.1.2}.  These `galaxies' have been rotated 
so that
each panel shows a projection along two of the principal axes; the
TREESPH $X-Y$ and $X-Z$ projections are shown on the left, and the 
corresponding GRAPESPH $X - Y$ and $X - Z$
projections are shown on the right. Halo 2 evidently forms a disc system
in both simulations. Similar numbers of stars are formed
with the two codes, but there are noticeable differences
attributable mainly to the much larger stellar softening 
length used in GRAPESPH. The TREESPH galaxy 
 is clearly  more concentrated, and has a much thinner
disc.

Much larger differences are evident in the evolution of the Halo 3
stellar systems, shown in the lower panels of Fig.~\ref{fig:fig5.1.2}.
The TREESPH galaxy shows
disc-like structure at both redshifts. At $z\approx 0.5$, the
main component in the GRAPESPH simulation is also disc-like,
but there is a satellite system containing about $25\%$ of 
the mass of the main galaxy which is not apparent in the TREESPH
panel. The satellite system is present in the TREESPH
simulation, but the slight difference in output redshifts ($0.49$ as opposed to
$0.46$ for GRAPESPH) coupled with the high relative speed of the intruder
conspire to move it outside the region plotted  in the figure.
In both runs the incoming satellite makes a first fly-by of the main disc
at $z \sim 0.4$, with the eventual merger occurring at $z \sim 0.15$.
However, because of the more extended structure of the GRAPESPH central
stellar disc that results from the larger softening, the interaction and
subsequent accretion of the satellite system is more disruptive in this case. 
As a consequence
the merger event has a preferentially destructive effect on the GRAPESPH
disc, leaving a spheroidal stellar object at $z=0$, whereas the TREESPH disc
survives largely intact. Referring back to Fig.~\ref{fig:fig5.1.1}, it is the 
subclump directly below the central clump that is
responsible for this collision (the systems to the left do not 
fall into the centre before $z=0$). This example shows the
sensitivity of the final remnant morphologies to the stellar softening. 

\subsubsection{Star Formation}

\begin{figure}
\centering
\centerline{\epsfxsize=8.0cm \epsfbox[110 220 480 600]
{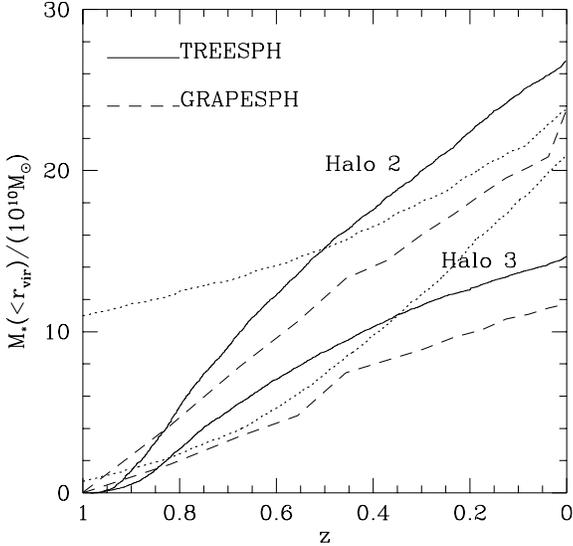}}
\caption{
Stellar mass within a virial radius for the Halo 2 and 3 galaxies between 
redshifts $z=1$ and $z=0$.  
Dotted lines are for TREESPH simulations of Halo 2
using the cooling suppression formula, $g(z)$ with $z_c=1$ (bottom line) and
with $z_c=4$ (top line).}
\label{fig:fig5.1.3c}
\end{figure}

We define a `virial' radius, $R_{vir}(z)$,
for each halo by placing a sphere at the
centre of mass within which the overdensity is equal to $\Delta =
178$.  The number of star particles in our simulations
within the virial radius at $z=0$
lies within the range $N_{star} = 1700 - 4000$.  In
Fig.~\ref{fig:fig5.1.3c}, the stellar mass within a virial radius for
the Halo 2 and 3 galaxies is shown over the redshift range $z=1$ to
$z=0$.  The two codes produce a comparable number of stars before
$z=0.8$, after which the TREESPH haloes accumulates stellar mass more
rapidly.  In both codes, stars are accumulated at a faster rate in
Halo 2 than in Halo 3.  The Halo 2 galaxy has a final stellar mass of
$2.4 \times 10^{11} M_{\odot}$ in GRAPESPH and $2.7 \times 10^{11}
M_{\odot}$ in TREESPH. For Halo 3, the final stellar mass is
$1.17\times 10^{11} M_{\odot}$ in the GRAPESPH run and $1.45 \times
10^{11} M_{\odot}$ in the TREESPH run.  Halo 2 forms the more massive
stellar disc, by about a factor of two.

It is interesting to compare these numbers with the stellar masses 
of  $L^*$ galaxies. The luminosity function
of Efstathiou, Ellis and Peterson (1988) gives 
$L^*_B = 5.1 \times 10^{10} L_\odot$
for $h = 0.5$. Assuming that the mass-to-light ratio of typical
discs is  $(M/L)_B \sim 3h$ in solar units,  the stellar mass
of a typical $L^*$ disc system is about $M*_{disc} \sim 8 \times 10^{10}
M_\odot$, \ie  somewhat smaller than the masses of the
stellar discs that form in Halos 2 and 3. The agreement is close 
enough, however, that it suggests that the basic physical picture of late
infalling gas may account for the masses of disc systems and the
origin of the Tully-Fisher relation (equation~\ref{ge7}).

Fig.~\ref{fig:fig5.1.3c} also shows the stellar mass within a virial radius 
for the
two TREESPH simulations of Halo 2 using the cooling suppression
formula, g(z) of  equation (\ref{gz}).  The bottom dotted line is for
a cut-off redshift of $z_c=1$ and the top dotted line is for $z_c=4$.
In both these simulations, radiative cooling and star formation are
turned on at $z=4$.  However, star formation is clearly nearly
suppressed in the $z_c=1$ model until $z=1$, after which stars
continue to form slowly  until $z=0$.
In the $z_c=4$ model, nearly half the final total mass
of stars has formed by $z=1$,  but by $z=0$ the galaxy
consists of slightly fewer stars than in the simulation with
cooling suppressed until $z_{\rm cool}=1$.

\begin{figure}
\centering
\centerline{\epsfxsize=8.0cm \epsfbox[80 180 400 600]
{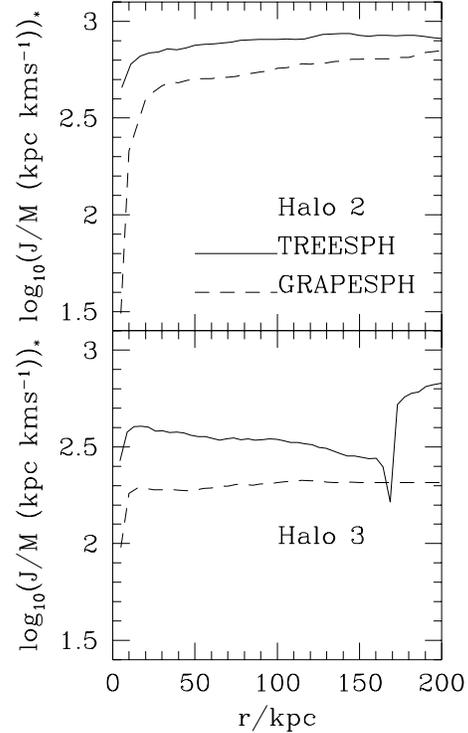}}
\caption{Cumulative specific angular momentum profile of the stellar
component out to 200 kpc for the Halo 2 and Halo 3 galaxies at $z=0$.}
\label{fig:fig5.1.4c}
\end{figure}

\subsubsection{Angular Momentum Profiles}

The cumulative specific angular momentum profiles of the stellar components in
Haloes 2 and 3 are shown in Fig.~\ref{fig:fig5.1.4c}.  
The total $J/M$ within a sphere is calculated for 90 equally-spaced radii.
The specific
angular momenta measured in the TREESPH simulations are about $50\%$
higher than in the GRAPESPH simulations.
A small companion at a distance $r \approx 170$ kpc appears as a
stepwise increase in specific angular momentum of Halo 3 in the
TREESPH run; this difference to the GRAPESPH run is a consequence
of the small differences in the dark matter evolution discussed
in Section~\ref{ssec:morph}. In Section~\ref{ssec:jevol}, 
we will show that angular momentum
evolution of the stellar and gaseous components can vary by an order
of magnitude or more, depending on what assumptions are made about
the physical parameters of the gas (\eg suppression of cooling).
These differences in $J/M$ arising from the input physics are therefore
much greater than the  differences between the codes seen in 
Fig.~\ref{fig:fig5.1.4c}.

The most obvious differences in Fig.~\ref{fig:fig5.1.4c} are in the
specific angular momenta in the central regions, 
$r \simlt 20$ kpc. On these small scales, the $J/M$
profile for the GRAPESPH runs declines much more
steeply than in the TREESPH runs. Evidently, the
angular momentum profiles decline rapidly on scales
smaller than a few softening lengths, and this effect
is much more pronounced in GRAPESPH because of the
much larger  stellar softening length. As described
in Section~\ref{ssec:morph}, the large softening length of
GRAPESPH leads to more diffuse stellar systems, and
clearly one cannot expect these simulations  to 
describe accurately the mass and velocity profiles
of real disc systems. However, global properties,
such as the total stellar mass and angular momentum
are much less sensitive to softening.

\subsubsection{Rotation velocities and velocity dispersions}

\begin{figure}
\centering
\centerline{\epsfxsize=8.0cm \epsfbox[100 180 500 600]
{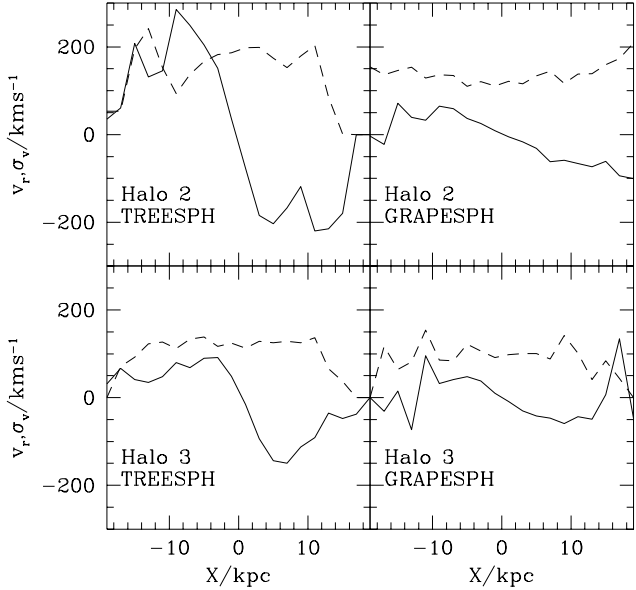}}
\caption{Projected major-axis velocity profiles out to $19$ kpc
for the Halo 2 and 3 stellar systems at $z=0$.  Solid lines show the
rotation velocity, $v_r$; dashed lines show the
projected velocity dispersion, $\sigma$. }
\label{fig:fig5.1.5c}
\end{figure}

Fig.~\ref{fig:fig5.1.5c} shows the projected velocity profiles out to
19 kpc of the Halo 2 and 3 galaxies. 
The projected velocities are calculated by distributing the particles
onto a Cartesian grid in which $X$, $Y$, and $Z$ are chosen to be
along the major, intermediate, and minor principal axes, respectively.
Each grid cell has dimensions of $\Delta l = 2$ kpc.
A slit with a width of two cells was laid parallel to the axis; results are
averaged over the two cells.
In Fig.~\ref{fig:fig5.1.5c}, the mean rotation speed of the stars
(solid lines) and the
projected velocity dispersion (dashed lines) are plotted against major axis 
distance
$X$, \ie as would be measured for an edge-on galaxy.

Clear evidence for systematic rotation is seen in both TREESPH runs. In 
the GRAPESPH runs, the Halo 2 galaxy is rotating with a 
peak amplitude of about $80\; \kms$. The difference with TREESPH
is, again, primarily a consequence of the larger GRAPESPH
softening length, which inhibits the formation of a compact, rapidly
rotating stellar system. Little rotation is seen in the GRAPESPH Halo 3 
stars, as might be expected 
from the lack of any disc in Fig.~\ref{fig:fig5.1.2}.  In comparison,
the TREESPH Halo 3 disc rotates with a maximum rotation
speed of  $v_r \approx 100$ km/s at 
$X \approx 3$ kpc. In both codes, the projected velocity dispersions
lie in the range $\sigma \approx 100 - 200$ km/s, and are
roughly constant with radius. The velocity
dispersions are comparable to the mean rotation speed in the TREESPH
code; thus, neither code succeeds in producing a cool, centrifugally
supported, stellar disc. This is almost certainly caused by two-body
heating (see \eg Steinmetz \& White 1997) which, for discs composed
of a few thousand particles, can heat a cold disc to
to $v_r/\sigma \sim 1$ within a few rotation periods. This is
consistent with the time-evolution of $\sigma$ which increases by
 a factor of $\approx 2$ between redshift $z=1$ and $0$.

\subsubsection{Stellar Mass Profiles}
\begin{figure}
\centering
\centerline{\epsfxsize=8.5cm \epsfbox[100 220 500 600]
{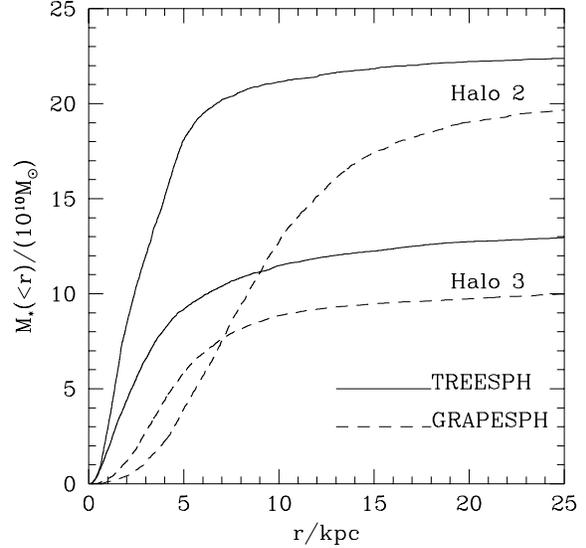}}
\caption{
Cumulative stellar mass profile of the Halo 2 and 3 galaxies
out to a radius of 25 kpc.}
\label{fig:fig5.1.6c}
\end{figure}

Fig.~\ref{fig:fig5.1.6c} shows the cumulative stellar mass profile 
for the Halo 2 and 3 galaxies out to a radius of 25 kpc.  
This gives an idea of the concentration of the final stellar
remnants and of the large differences caused by the softening
in GRAPESPH. Even in  the TREESPH code, the half-mass radii
of the stellar systems are $\sim 3$ kpc, about equal to
stellar softening. This shows that it is possible to form
rapidly rotating discs with similar scale lengths to real
disc galaxies, though higher resolution calculations are
required to model their mass and velocity profiles.

\subsubsection{Summary}
The simulation of disc galaxy formation, including star
formation, is a complex problem and it is often difficult to
assess the accuracy of any particular computational technique.
The purpose of this subsection has been to compare the results
of evolving identical conditions with two independent SPH codes
with the aim of finding which properties, if any,  are sensitive to numerical
methods. To summarise the findings, the dark matter and gas 
evolution is broadly similar. However, in one of the $5$ haloes 
(Halo 3), the small differences that do exist 
produce a significantly different stellar remnant. The overall
morphologies of the stellar systems that form
in the other four haloes are more similar.
Although the star formation algorithms are different in the
two codes, they  give similar star formation
rates and final stellar masses. This is not too surprising because
the algorithms are designed to convert high density cool gas into stars 
on a timescale of order the dynamical time, which is much shorter
than the Hubble time when most of the stars form in these simulations.
The star formation rate is therefore governed mainly by the rate
at which gas infalls  towards the halo centres, and this
is modeled in a similar way in both codes.

The main differences between the codes are attributable to the larger
stellar softening employed in the GRAPESPH simulations. This results
in stellar systems that are more extended and have lower rotation
velocities than those that form in the TREESPH simulations. Even with
TREESPH, the stellar softening of $2.6$ kpc is barely adequate to
resolve the internal structure of the discs that form. 
Despite these differences, the global properties of the stellar
remnants, \eg their masses and angular momenta, are in 
reasonably good agreement. In the next Section we will investigate,
using both codes, how these global properties depend on the 
input physics.

\subsection{Angular Momentum Evolution}\label{ssec:jevol}

\begin{figure}
\centering
\centerline{\epsfxsize=8.0cm \epsfbox[30 170 400 700]
{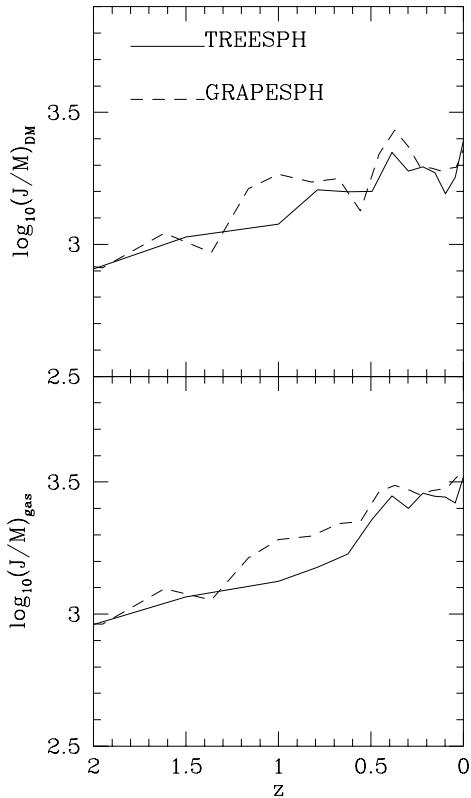}}
\caption{Evolution of the mean specific angular momentum within a virial
radius of the dark matter (top) and gas (bottom) of the five simulations
run with $z_{\rm cool}=1$.  Here and in all succeeding figures, $J/M$ is in 
kpc km s$^{-1}$.}
\label{fig:fig5.2.1c}
\end{figure}

\begin{figure*}
\centering
\centerline{\epsfxsize=6.0cm \epsfbox[200 400 390 650]
{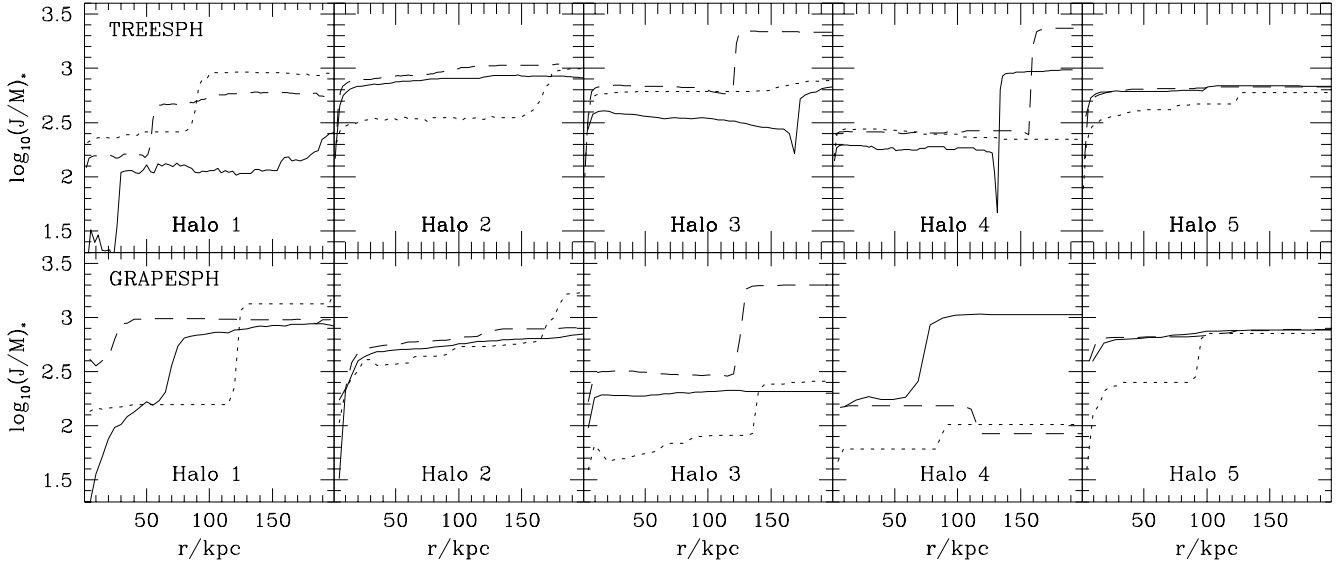}}
\caption{Radial specific angular momentum profiles of the stellar components 
in each halo for the simulations with $z_{\rm cool}=1$.  
Redshifts for TREESPH results are 
$z=0.63$ (dotted lines), $0.30$ (dashed lines), and $0.0$ (solid lines); 
redshifts for GRAPESPH results are similar except that the first redshift 
is $z=0.68$.}
\label{fig:fig5.2.2c}
\end{figure*}

We first discuss the evolution of angular momentum in the simulations
in which cooling is suppressed until $z_{\rm cool}=1$. We then discuss how
varying $z_{\rm cool}$ alters the conclusions.

The accumulation of angular momentum by dark matter haloes and their
associated gas proceeds smoothly during the expansion phase before
turnaround, growing roughly as $t^{5/3}$ (White 1984, Barnes and
Efstathiou 1987).  After collapse, halo angular momentum growth
slows and then evolves by abrupt jumps caused by  discrete merger
events.  
Since the haloes are all of a
similar size, their total specific angular momentum within 
the virial radius, $R_{vir}(z)$,
was averaged in order to give a statistical
representation of the growth of $J/M$ with redshift.
The evolutions of the mean specific angular momentum for the gas
and dark matter are shown in Fig.~\ref{fig:fig5.2.1c}.
The TREESPH  and GRAPESPH results are plotted as solid and
dashed lines respectively, and are almost identical. 
The specific angular momenta of the dark matter and gas have 
very similar values, but at late times the specific angular
momentum of the gas component exceeds that of the dark matter.
This is because  $J/M$ of both the components increases 
with radius. Consequently, when
the central, low $J/M$, gas turns into stars, the specific angular momentum
of the remaining gas within the virial radius is enhanced.

Fig.~\ref{fig:fig5.2.2c} shows the radial specific angular momentum
profiles of the stellar components in each halo for both TREESPH 
and GRAPESPH simulations at three different epochs. 
The redshifts for TREESPH results are $z=0.63$ (dotted lines), $0.30$ 
(dashed lines), and $0.0$ (solid lines); for GRAPESPH, the first redshift is
$z=0.68$.
It can be inferred  from the large number of almost horizontal
lines with the occasional vertical jump, that the dominant
stellar concentrations  are usually
well isolated from other star particles. 
The
two most disc-like final stellar components, belonging to Haloes 2 and 5, 
evolve relatively steadily from $z=0.6$ to the present, whilst
the remaining haloes suffer more interactions with other stellar clumps.
The differences  in the positions and sizes of the vertical jumps in $J/M$ in
identical halos run with TREESPH and  GRAPESPH are caused
by the evolutionary differences identified in Section 5.2.1 that 
lead to sub-condensations in each code with different
orbits.

\begin{figure}
\centering
\centerline{\epsfxsize=8.0cm \epsfbox[100 200 500 700]
{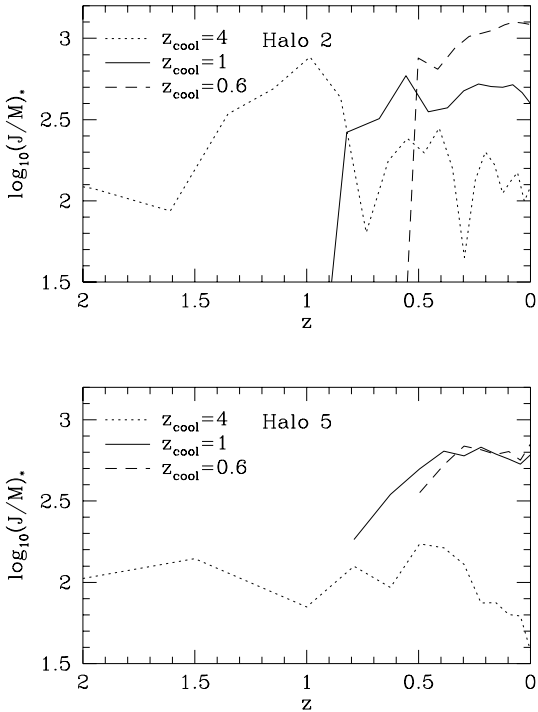}}
\caption{Evolution of the stellar $J/M$ measured within the central $20$kpc 
for GRAPESPH Halo 2 and TREESPH Halo 5.}
\label{fig:fig5.2.3}
\end{figure}

The morphology and specific angular momenta of the final stellar
objects  appear to be sensitive to the merger histories undergone
by the parent haloes. This point has been investigated quantitatively
by resimulating two haloes with the cooling switched on at
$z_{\rm cool}=4$ and $z_{\rm cool}=0.6$, in addition to the usual
$z_{\rm cool}=1$.  The evolution of the stellar $J/M$ measured within the
central $20$ kpc is shown in Fig.~\ref{fig:fig5.2.3}. The top frame
shows GRAPESPH results for Halo $2$; the bottom frame shows TREESPH
results for Halo $5$.  For Halo $2$, it is clear that a significant
increase in the final specific angular momentum occurs when the gas is
prevented from cooling at early times. The same result is reproduced
by the Halo $5$ simulations, but only for the change of $z_{\rm cool}$
from $4$ to $1$. This halo evolves very little after $z=1$ (see Table
1); the result of suppressing cooling until $z_{\rm cool}=0.6$ is to form
a very similar stellar disc as in the $z_{\rm cool}=1$ simulation, but at
a slightly later time.  From $z=1$ to the present, Halo $5$ increases
its virial mass by a factor of $\sim 1.5$ whereas Halo $2$ grows by a
factor of $\sim 1.2$. The other three haloes nearly double in mass
during this period.

\begin{figure*}
\centering
\centerline{\epsfxsize=15.0cm \epsfbox
{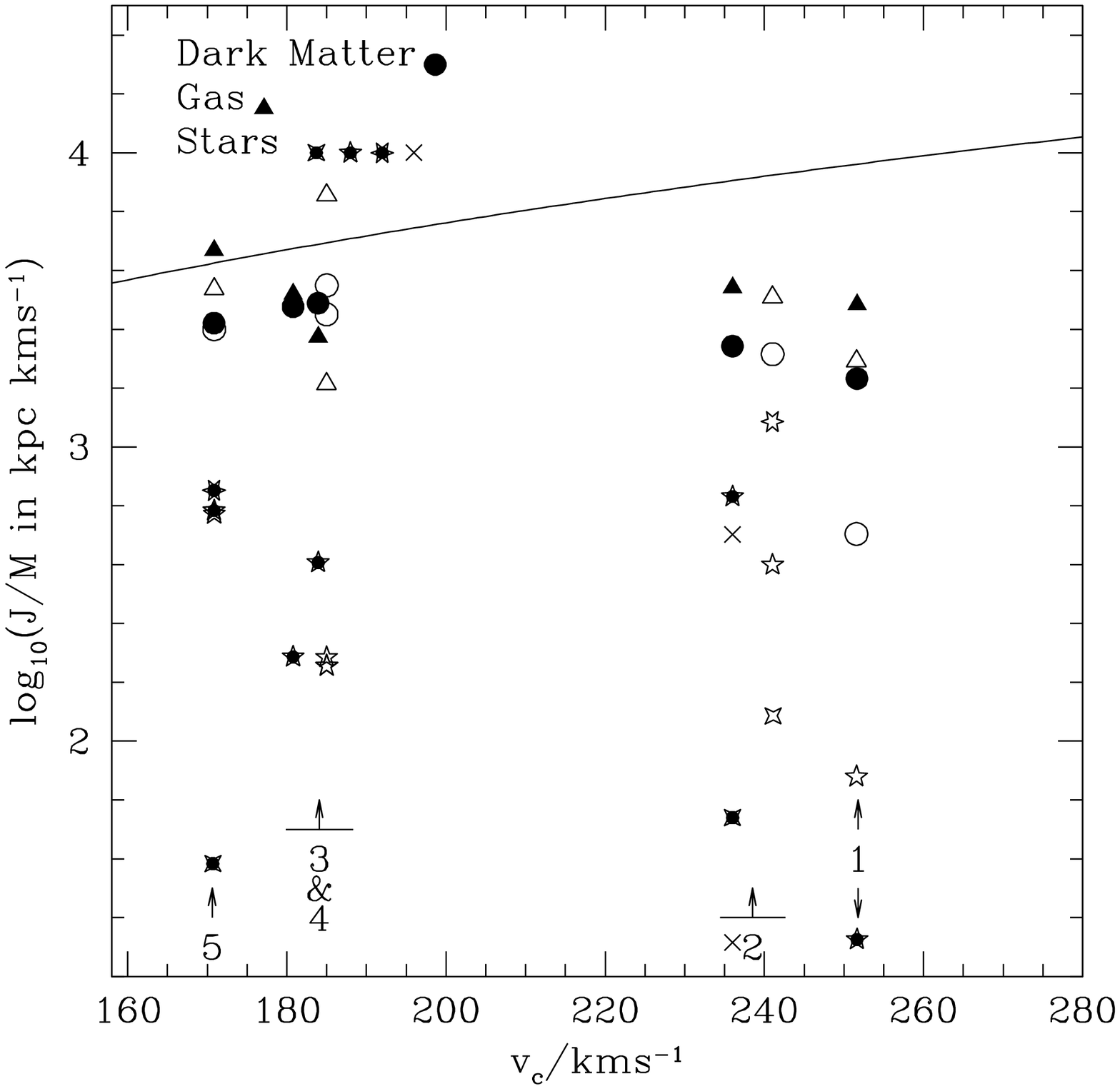}}
\caption{Specific angular momenta at $z=0$ versus circular
speed $v_c$ for stars within $20\;{\rm kpc}$ (plotted as stars) 
gas (plotted as triangles) and dark matter (plotted as circles) 
at the virial radius. TREESPH results are shown by the 
filled symbols and GRAPESPH results by the open symbols.  For the stellar
components, 
$4-$pointed stars show results for runs with $z_{\rm cool}=4$, $5-$pointed
stars for $z_{\rm cool}=1$, and 
$6-$pointed stars  for $z_{\rm cool}=0.6$.  
Crosses are for Halo 2 stars evolved
with cooling suppressed by $g(z)$.  The abscissa is the circular velocity 
of the halo at the viral radius. The solid line is equation~\ref{ge5}.}
\label{fig:fig5.2.4c2}
\end{figure*}

Fig.~\ref{fig:fig5.2.4c2} shows the specific angular
momenta of all the three components (stars at $20$ kpc, gas and dark matter at
the virial radius) for both TREESPH (filled symbols) and GRAPESPH (open
symbols) haloes at redshift zero. 
The circular speed of the haloes is computed from equation (1) with
$R_c$ set to the virial radius $R_{vir}$.
Concentrating on the stellar components (plotted as $4-$pointed stars for
$z_{\rm cool}=4$, $5-$pointed stars for $z_{\rm cool}=1$ and $6-$pointed stars
for $z_{\rm cool}=0.6$),
it can be seen that, even allowing for the differences between the two codes,
there is a tendency for the more slowly evolving haloes 
(numbers $2$ and $5$) to give rise to more rapidly rotating 
stellar objects, {\it provided cooling is suppressed at high
redshift}. If cooling is allowed to begin at
$z_{\rm cool}=4$, the stellar systems that form in these haloes
experience the `angular-momentum catastrophe' seen in previous
numerical simulations.
The crosses in  Fig.~\ref{fig:fig5.2.4c2} show 
the specific angular momenta of the stars for the 
Halo 2 TREESPH simulations in which cooling was suppressed
using the  empirical formula $g(z)$ of
equation (\ref{gz}).  The lower cross is for $z_c=4$ and the upper cross is 
for $z_c=1$.  The specific angular momentum of the $z_c=1$ simulation
is an order of magnitude larger than the $z_c=4$ run, though it is not
as high as in the simulation with cooling abruptly truncated at $z_{\rm cool}=1$.
Nevertheless, these simulations demonstrate again the sensitivity of
angular momentum evolution to the cooling history.

The solid line in Fig.~\ref{fig:fig5.2.4c2} shows equation~\ref{ge5}.
The specific angular momenta of the dark matter haloes all lie below
this line. This is expected because, as described in Section 4.1, our
selection criteria bias against haloes with final spin-parameters
larger than $\lambda = 0.05$. All of our haloes therefore have spin
parameters less than or equal to 
the median value of $\lambda$ measured in N-body
simulations (see Figure 1) and hence lie below the line plotted in
Fig.~\ref{fig:fig5.2.4c2}.

The gas components generally have higher final specific angular
momenta than their dark matter haloes. As explained at the start of
this subsection, this is because the lower-angular momentum gas in the
central regions of the haloes is efficiently converted into stars,
leaving an extended atmosphere of high angular momentum gas.

There is good agreement between the TREESPH and GRAPESPH runs
plotted in Fig.~\ref{fig:fig5.2.4c2}, with one exception, namely
the final specific angular momentum of the stellar components
of the halo 3 runs with $z_{\rm cool}=1$ (see Section 5.2.1, for a 
detailed discussion).  The general  conclusion
from both the  TREESPH and GRAPESPH simulations is the same however:
the stellar systems that form in simulations with high redshift
cooling experience an `angular-momentum catastrophe' and end
up with specific angular momenta that are between one or two orders
of magnitude smaller than the specific angular momenta of real
disc galaxies. In contrast, if cooling is suppressed until
$z_{\rm cool} \sim 1$, stellar discs can form with angular momenta
that are within the range observed for real galaxies. For example,
comparing the $z_{\rm cool}=1$ results plotted in 
Fig.~\ref{fig:fig5.2.4c2} with the results of 
Figure 1 shows that the stellar systems that form in haloes
5, 3 (in the TREESPH run) and $2$ have similar specific
angular momenta to real disc galaxies. They also have smaller
specific angular momenta, by a factor of about 4, compared to
their parent dark matter haloes at $z=0$, in agreement with the
results plotted in Figure 1. These results provide powerful
evidence that late disc formation is essential if we are
to explain their angular momenta.

However, suppressing cooling until $z_{\rm cool}=1$ does not guarantee
that disc systems will form by the present day. Thus, even with
cooling suppressed until such a recent epoch, the stellar remnants
that form in Haloes 1,  4 (and Halo 3 in the GRAPESPH run)
have much lower specific angular momenta
than real disc galaxies. This is because these dark matter haloes 
evolve considerably between $z=1$ and the present (nearly doubling
their masses within the virial radius); star formation in these haloes
occurs in sub-units which merge between $z=1$ and the present day,
losing angular momentum in the process. The formation of discs thus
depends sensitively on the internal structure and  merger history 
of their parent dark matter haloes at recent epochs.

\subsection{Colour Evolution}

\begin{figure}
\centering
\centerline{\epsfxsize=8.0cm \epsfbox[60 200 400 690]
{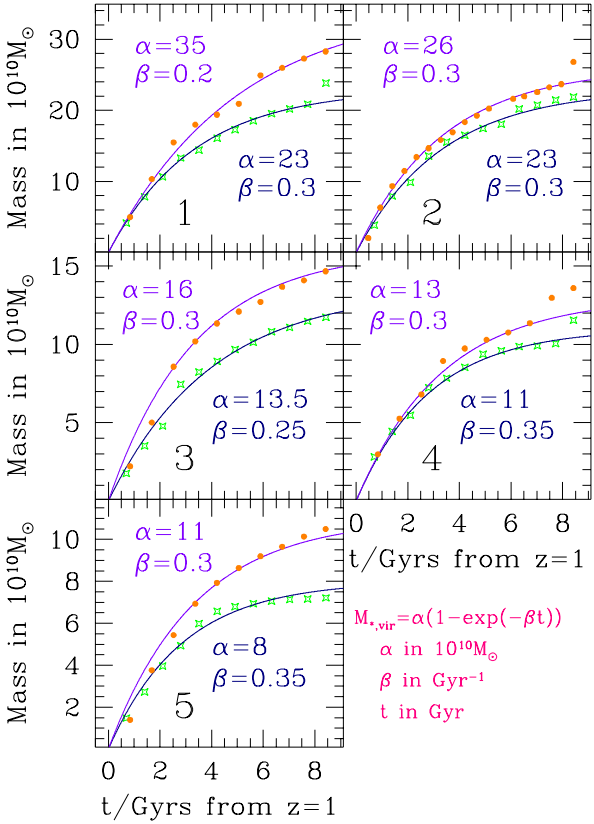}}
\caption{Star formation histories for the haloes evolved with $z_{\rm cool}=1$.
The TREESPH runs are shown as the filled circles and the GRAPESPH
runs are shown as the open  stars. The halo numbers are listed in
each panel. The solid lines show fits of equation
\ref{vre1} with values of $\alpha$ and $\beta$ as given in each panel.}
\label{fig:fig15}
\end{figure}

The star formation histories for the five models with $z_{\rm cool}=1$
are shown in   Fig.~\ref{fig:fig15}. As explained in Section 5.2.2, the
star formation histories in the TREESPH and GRAPESPH runs are slightly
different, with more net star formation in the TREESPH models. However,
in both sets of models, the star-formation histories are quite well
approximated by the form
\begin{equation}
 M_*(t^\prime) = \alpha ( 1 - {\rm exp}\;(-\beta t^\prime)), \label{vre1}
\end{equation}
with values of $\alpha$ and $\beta$ as listed in Fig.~\ref{fig:fig15}.
In equation \ref{vre1}, $t^\prime$ denotes the time measured from $z=1$
when cooling is switched on. In these models, there is 
no star formation by construction at $z > 1$, and so the entire 
present day galaxy forms since $z=1$. The star formation rates decline
once star formation begins. In the extreme
cases shown in Fig.~\ref{fig:fig15}, the star formation rates vary
between $\dot M \approx 70 M_\odot/{\rm yr}$ at $z=1$ to 
$\dot M\approx  12M_\odot/{\rm yr}$ at $z=0$ for Halo 1 
(which does not form a final disc) and  
$\dot M \approx 28 M_\odot/{\rm yr}$ at $z=1$ and 
$\dot M\approx  1.5 M_\odot/{\rm yr}$ at $z=0$ for Halo 5
(which does form a stellar disc).

\begin{figure}
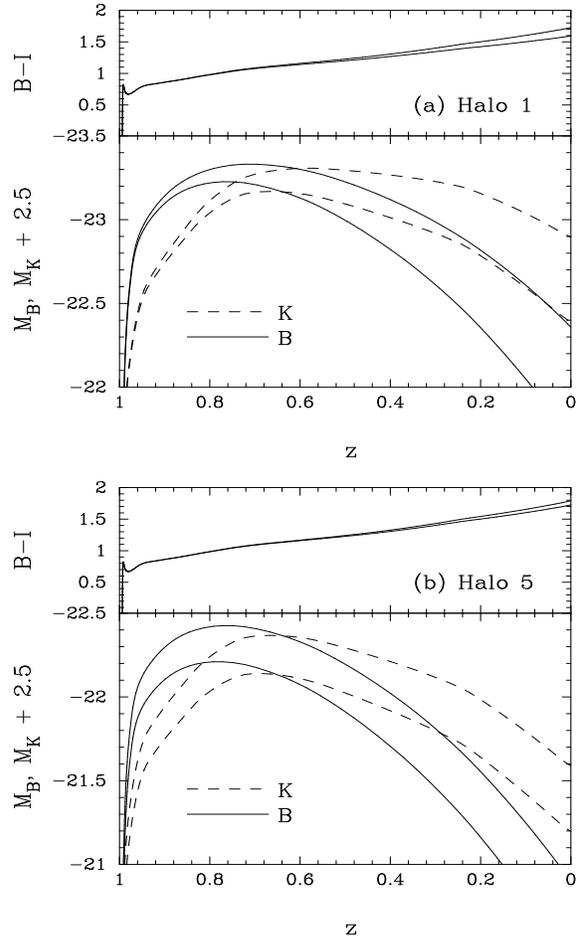


\vskip 5.2 truein

\includegraphics{pg_6.2a.ps}
\includegraphics{pg_6.2b.ps}

\caption{Bruzual and Charlot models of the rest-frame
absolute magnitude evolution in the $B$ and $K$ bands, 
and $B-I$ colours, for the star formation rates in Haloes
$1$ and $5$ as plotted in Figure \ref{fig:fig15}. The pairs of
lines show the evolution determined for the
star formation rates from the TREESPH and GRAPESPH 
simulations. A standard Salpeter (1955) stellar initial mass
function was assumed. }
\label{fig16}
\end{figure}

In the previous subsection, we have argued that high angular
momenta of disc galaxies requires that they formed at late 
epochs. Thus, the question  arises as to whether 
such extreme evolution as implied by Fig.~\ref{fig:fig15}
could be compatible with observations of disc galaxies 
at redshifts $z \sim 0.5$--$1$ (Lilly \etal 1997, Vogt \etal
1997a, b). We have therefore computed the rest frame
absolute magnitudes and colours for Haloes 1 and 5
(spanning the range of star formation histories plotted
in Fig.~\ref{fig:fig15}) using the Bruzual and Charlot
(1993) population synthesis code. The results are plotted
in Fig.~\ref{fig16} for a Salpeter (1955) initial stellar 
mass function extending from $0.1M_\odot$ to $125 M_\odot$. These
models show that there is weak evolution in the K-band, once
star formation is underway. The rest-frame B-band absolute
magnitudes brighten by more than a magnitude between $z=0$ and
$z=0.8$, despite the fact that the stellar disc mass decreases 
by an order of magnitude between these epochs. This is not surprising
because the luminosity, especially in the B-band,
is sensitive to the net star-formation rate at early times rather than the
underlying disc mass. 

An analysis of recent observations of disc evolution (Mao, Mo and
White 1998), suggests that at fixed circular speed, discs show very
little evolution in B-band luminosity. Comparison with
Fig. \ref{fig16} suggests that the star formation rates in our
simulations are probably too high at early times. In the simple
cooling scheme adopted in our models, a reservoir of high density gas
builds up in the central regions of the haloes that cools within
a dynamical time and forms stars once cooling is switched on.  It is
therefore likely that the star formation rates will be less strongly
peaked to high redshifts in more realistic feedback models and hence
that the luminosity evolution will be even weaker than that shown in
Fig. \ref{fig16}. 

The very strong evolution of disc masses with redshift should manifest
itself as a strong evolution of the disc structural parameters (scale
lengths) with redshift.  Discs at high redshift should have smaller
scale lengths than at the present day. However, the gravitational 
softenings in our
simulations are comparable to the disc scale lengths and so we cannot
reliably predict either their size or how they will evolve with redshift. 
(In most of our runs, the stellar disc half-mass radii are about equal to
the stellar softening and stay roughly constant as the discs evolve.)
Nevertheless, Mao \etal (1998) find evidence that at fixed circular speed, disc
scale lengths do decrease as $(1+z)^{-1}$. This is compatible both with simple
analytical models of disc evolution (\eg Mo \etal 1998) and
with the conclusions in favour of late disc formation presented
in this paper.

\section{Conclusions}\label{sec:conc}

The purpose of this study has been to investigate whether disc
galaxies can form in a CDM universe with angular momenta and sizes
similar to those of real disc systems. We have computed the specific
angular momenta of CDM haloes from an N-body simulation and compared
them with estimates for a large sample of disc galaxies.  This
comparison shows that disc galaxies have very nearly the same
specific angular momenta as those  within the virialized
region of CDM haloes at $z\approx 1$.  We therefore conclude that in a
CDM-like model, the gas that formed the disc component must have
collapsed at recent epochs,  very nearly conserving its angular
momentum during collapse. This picture is capable of explaining the
present day sizes and distribution of surface brightnesses of 
disc galaxies (see also Dalcanton \etal 1997, 
and Mo \etal 1998).

However, numerical simulations of disc formation have shown that the
gas experiences an `angular momentum catastrophe' if it is allowed to 
collapse early
(\eg Navarro \& Benz, 1991; Navarro, Frenk \& White, 1995a; 
Navarro \& Steinmetz 1997). In the absence of feedback
processes, the gas collapses into subunits at high redshift, which
subsequently merge losing most of their angular momentum to the dark
halo material.  The specific angular momenta of the resulting gaseous
discs are typically two orders of magnitude smaller than those of real
disc galaxies.

In this paper, we have tested whether the `angular momentum
catastrophe' can be avoided if feedback processes prevent the gas from
collapsing until late epochs when haloes are reasonably smooth and
slowly evolving. If this can be achieved, it may be possible for the
gas to conserve most of its angular momentum during disc formation, as
assumed in the simple analytical models of Fall and Efstathiou (1980)
and others.  We have not attempted to model feedback in any detail in
this paper, but have instead suppressed cooling until a specified
redshift. We have selected five haloes with present day circular
speeds in the range $150 \simlt v_c \simlt 270$ km/s from a
dissipationless N-body simulation of an $\Omega=1$ CDM model.
We have generated higher resolution multi-mass  initial conditions 
of these haloes which we have evolved with two SPH codes,
TREESPH and GRAPESPH, that include star formation.

The results from the two codes are generally in good agreement.  Most
of the differences are caused by the larger softening length of the
stellar component of the GRAPESPH code which affects the internal
structure of the collapsed objects and their susceptibility to
disruption by mergers. In one pair of simulations (runs 11 and 12) the
final stellar system has a different morphological appearance in the
two codes, appearing disc-like in TREESPH and spheroidal in
GRAPESPH. However, in all other simulations, the morphologies of the
final stellar systems are qualitatively similar in the two codes.

All of the runs with cooling suppressed until $z_{\rm cool}=4$ produced
spheroidal systems with final  specific angular momenta of between
one and two orders of magnitude lower than their  dark matter
haloes. These systems experience the `angular momentum catastrophe'
seen in previous simulations. However, in the simulations
with  cooling  suppressed until $z_{\rm cool}=1$, stellar discs can
form with specific angular momenta similar to those of real disc
galaxies. This occurs in three out of five haloes simulated with TREESPH
and two out of five simulated with GRAPESPH. Suppressing disc formation
until late epochs can therefore solve the `angular momentum catastrophe'
in some cases, but does not guarantee that a disc will survive to the
present day. The  morphology of the final stellar object in the
simulations depends sensitively on the merging history of
the parent halo. All of the haloes that we have simulated have been
chosen so that they do not merge with a comparable mass system at
$z \simlt 1$. Nevertheless, Haloes  1 and 4, approximately double
their mass between $z=1$ and $z=0$ through a succession of 
mergers and do not maintain stellar discs to the present day. Haloes
$2$ and $5$ grow least between $z=1$ and $z=0$ (see Table 1) and 
these produce the most convincing stellar discs in our simulations.

Our simulations have demonstrated that angular momentum evolution
during galaxy formation is extremely sensitive to the thermal history
of the gas. This suggests that feedback processes can solve the
`angular momentum catastrophe' and are a necessary ingredient in 
disc galaxy formation. The structure of dark matter haloes is also
important in determining the angular momentum evolution. In fact, a number
of authors (Katz \& Gunn, 1991, Vedel \etal, 1994, 
Steinmetz \& Muller, 1995, Contardo, Steinmetz and Fritze-von
Alvensleben 1998) have shown that gas approximately conserves
its angular momentum if it collapses within nearly 
uniform dark matter haloes that grow by monolithic collapse rather
than hierarchical merging. This is not a realistic solution, however,
within the context of CDM-like theories. The detailed evolution of
dark matter haloes in CDM models will depend on the cosmology and
spectrum of fluctuations and these may be important in determining
the fraction of disc galaxies at the present day. To address this issue in 
an unbiased fashion would require a random choice of 
haloes to be resimulated, unlike that adopted here.

Our simulations suggest that the disc components of galaxies formed at
recent epochs, perhaps as late as $z \simlt 1$. This is in qualitative
agreement with recent observations showing that discs are evolving
significantly over this redshift range (Mao \etal 1998). A more
detailed comparison with observations will, however, require higher
resolution simulations and more realistic models of feedback and 
star formation.

\section*{ACKNOWLEDGMENTS}
We thank Yong-Ik Byun and Ken Freeman for allowing us to 
use their disc photometry parameters prior to publication.
MLW acknowledges funding by a PPARC postdoctoral position  at the
University of Oxford. VRE acknowledges Douglas Heggie for 
looking after the GRAPE in Edinburgh and 
the support of a PPARC postdoctoral fellowship. GPE thanks
PPARC for the award of a Senior Fellowship.

\end {document}